\documentclass[journal]{IEEEtai}

\usepackage[colorlinks,urlcolor=blue,linkcolor=blue,citecolor=blue]{hyperref}

\usepackage{color,array}

\usepackage{graphicx}
\usepackage{amsmath,amsfonts}
\usepackage{algorithm}
\usepackage{algorithmicx}
\usepackage{algpseudocode}
\usepackage{booktabs}
\usepackage{amsfonts}
\usepackage{nicefrac}
\usepackage{microtype}
\usepackage{array}
\usepackage[caption=false,font=normalsize,labelfont=sf,textfont=sf]{subfig}
\usepackage{textcomp}
\usepackage{xcolor}
\usepackage{stfloats}
\usepackage{url}
\usepackage{amsthm}
\usepackage{amsmath}
\usepackage{amssymb}
\usepackage{multirow}
\usepackage{mdframed}
\usepackage{makecell}
\usepackage{tabularx}
\newtheorem{theorem}{Theorem}
\newtheorem{lemma}{Lemma}

\theoremstyle{definition}
\newtheorem{definition}[theorem]{Definition}

\theoremstyle{remark}
\newtheorem{remark}[theorem]{Remark}
\begin{document}

\title{ConfusionPrompt: Practical Private Inference for Online Large Language Models} 

\author{Peihua Mai, Ran Yan, Youjia Yang, Rui Ye, Wei Liu, Yan Pang
\thanks{Corresponding author: Yan Pang (e-mail: jamespang@nus.edu.sg).}}

\maketitle

\begin{abstract}
State-of-the-art large language models (LLM) are often deployed as online services, posing privacy risks for user prompts. In response, we introduce ConfusionPrompt, a novel framework for private LLM inference that protects user privacy by: (i) decomposing the original prompt into smaller sub-prompts, and (ii) generating pseudo-prompts alongside the genuine sub-prompts. ConfusionPrompt seamlessly integrates with existing black-box LLMs and achieves a superior privacy-utility trade-off compared to existing text perturbation methods. We develop a $(\lambda, \mu, \rho)$-privacy model to formulate the requirements for a privacy-preserving group of prompts and provide a complexity analysis to justify the role of prompt decomposition. Our empirical evaluation shows that ConfusionPrompt outperforms local inference with open-source models by over 11\% and surpasses perturbation-based techniques by over 42\%, while reducing memory consumption by at least 62\% compared to open-source LLMs.
\end{abstract}

\begin{IEEEImpStatement}
The rapid adoption of LLMs via cloud-based services has heightened privacy concerns, as users must transmit sensitive prompts to external servers. Existing approaches for privacy-preserving LLM inference, such as encryption-based techniques and perturbation-based strategies, are often impractical as they either significantly degrade model performance or lack compatibility with black-box LLMs. Our proposed framework, ConfusionPrompt, introduces a novel, client-side framework that enables private inference without requiring any modifications to existing black-box LLM services. ConfusionPrompt consistently outperforms local inference with open-source LLMs and perturbation-based privacy methods in terms of utility, while also minimizing memory and computational overhead. This framework advances toward safer, privacy-conscious deployment of LLMs, an increasingly pivotal concern in real-world AI applications.
\end{IEEEImpStatement}


\section{Introduction}
\label{sec:intro}
Large language models (LLMs) have demonstrated remarkable capabilities across a wide range of tasks \cite{brown2020language, roziere2023code, huang2023instruct2act, chen2025improving,wu2025ensemble}, driving their adoption in real-world applications such as medical consultations and financial services \cite{wu2023bloomberggpt, wang2025imagent,thirunavukarasu2023large}. State-of-the-art LLMs, such as OpenAI's ChatGPT \cite{ouyang2022training}, are predominantly offered as online services, primarily to safeguard proprietary model parameters and intellectual property \cite{openai2023gpt4}. However, this paradigm introduces significant privacy risks, as users often transmit prompts containing sensitive information that should ideally remain confidential from the server.

Existing solutions to privacy-preserving model inference, such as encryption \cite{liu2023llms, chen2022x} and perturbation \cite{du2023dp} techniques, are often impractical for private LLM inference. In encryption-based methods, the server leverages homomorphic encryption (HE) \cite{acar2018survey} and Secure Multiparty Computation (SMPC) \cite{cramer2015secure} to conduct private inference on the users’ encrypted query.
The practicality of this method is limited by its high computation overheads, especially on LLMs (e.g., GPT-3 has 175B parameters \cite{brown2020language}). On the other hand, the basic idea of perturbation-based methods is to inject a specific level of noise into the user's input before releasing it to the server, which is challenging to strike a satisfied privacy-utility balance \cite{lyu2020differentially}.

Another limitation of the aforementioned solutions is their dependence on the service provider to modify their infrastructure or disclose specific model parameters, which can be prohibitively costly. Encryption-based techniques require significant investment from platform in hardware and algorithmic acceleration to support efficient computation. Perturbation-based methods often rely on local differential privacy (LDP) \cite{dwork2006differential} to ensure privacy with formal guarantees. For improved LDP performance, recent studies have proposed deploying certain modules on the user side, necessitating the sharing of specific model parameters by the server \cite{qu2021natural, maisplit, malekzadeh2024salted, li2024data}. These requirements impose additional burdens and proprietary concerns on the service provider, hindering the deployment of these solutions. It remains an open challenge to \emph{develop a method that seamlessly integrates with existing black-box LLMs while maintaining strong utility}.

To alleviate the above concerns, this paper proposes ConfusionPrompt, a novel private LLM inference framework that can be seamlessly integrated with existing online black-box LLMs.
The core idea of ConfusionPrompt is to construct a set of prompts, containing real and fake prompts, designed to confuse the server, preventing the server from accurately identifying the genuine user prompt. However, a significant challenge arises when a prompt contains multiple sensitive attributes. In such cases, the user needs to consider all combinations of true and fake attributes to effectively confuse an attacker with prior knowledge of certain attributes, rendering the query complexity growing exponentially with the number of attributes. To address this issue, we propose a decomposition approach in which the query is divided into sub-questions. This ensures that the private information is distributed across different sub-queries, thereby reducing the complexity to approximately linear in the number of attributes (see Figure \ref{fig:decompexamreal}). 

Building on the aforementioned concepts, our proposed ConfusionPrompt consists of four critical steps:
(1) The user decomposes the original prompt into several sub-prompts.
(2) For each genuine sub-prompt, the user generates a set of pseudo prompts that obfuscate sensitive attributes in the genuine sub-prompt, repeating this process until the desired level of privacy is achieved.
(3) The user sends a group of both genuine and pseudo-prompts to the online service, retrieves the responses, and isolates the sub-responses corresponding to the genuine sub-prompts.
(4) Finally, user recomposes the sub-responses to obtain the final result.
This framework allows for independent user-side deployment, eliminating the need for privacy-preserving modifications by the service provider.

Accordingly, we develop a $(\lambda, \mu, \rho)$-privacy model for ConfusionPrompt to formulate the requirements for a privacy-preserving group of prompts.
Through complexity analysis, we show that for a prompt $\boldsymbol{p}$ with $U(\boldsymbol{p})$ private attributes and a privacy budget $\mu$, the basic confusion strategy without decomposition requires generating $\mathcal{O}\left((1/ \mu)^{U(\boldsymbol{p})}\right)$ fake prompts,
while our decomposition strategy in ConfusionPrompt can reduce the complexity to $\mathcal{O}\left( (1/\mu) U(\boldsymbol{p}) \right)$ in the ideal decomposition scenario.
Based on the privacy model and complexity analysis, we derive the criteria for an ideal decomposer and generator, and accordingly design a two-stage training strategy.

Our key contributions are as follows:

(1) We are the first to propose a private LLM inference framework using confusion-based strategy. Our ConfusionPrompt framework can be seamlessly implemented by clients with existing online black-box LLMs, such as ChatGPT and Claude, providing an improved privacy-utility trade-off. In addition, our framework can be integrated with Trusted Execution Environment (TEE) \cite{sabt2015trusted} solutions. Even in the event of a TEE compromise, user privacy remains safeguarded through the confusion strategy.

(2) To ensure user privacy, we define a privacy model to formulate the requirements for a group of prompts, including both real and pseudo-prompts. Furthermore, we introduce a local decomposition module to reduce the complexity of the prompt group under the same privacy parameters.

(3) Experiments show that ConfusionPrompt can achieve a consistently and significantly better utility than LDP-based methods. Moreover, ConfusionPrompt surpasses the performance of local inference method using open-source models with lower memory requirement.

\section{Related Works}
\label{sec:ltr}
In this section, we review three types of methods for privacy-preserving LLM inference: sanitization-based method, encryption-based method, and perturbation-based method.

\textit{Sanitization-based method.} Traditional sanitization techniques rely on Named Entity Recognition (NER) to redact Personally Identifiable Information (PII) such as individuals’ names, social security number, and email addresses \cite{ehrmann2023named, lample2016neural, qiao2026beyond}. The downside of this method is that NER falls short in concealing other sensitive information, including verbs and non-named entities \cite{anandan2012t}. Recent research on LLM inference involves sanitizing sensitive items in the input and subsequently de-anonymize the LLM’s returned responses \cite{kan2023protecting, chen2023hide,qiaohessian}. However, such input perturbations can introduce semantic biases, potentially leading to responses that are semantically irrelevant. Casper utilizes a combination of rule-based and machine learning (ML)-based techniques to filter sensitive PIIs and alert users of the sensitive topics \cite{chong2024casper}. While this method effectively detects sensitive information, it does not address how to process such information without adversely affecting the output of LLMs. In general, sanitization approaches identify and remove sensitive information, but this can degrade output utility when such information is crucial for LLM inference.

\textit{Encryption-based method.} Cryptonets \cite{gilad2016cryptonets} proposed the first neural network inference on encrypted data using homomorphic encryption (HE). They approximated the non-linear function such as Sigmoid and MaxPooling by polynomials. Iron \cite{hao2022iron} designed specialized and efficient protocols for two types of computationally heavy operations in Transformer-based inference: (i) matrix multiplications, and (ii) complex functions including Softmax, GELU activations, and LayerNorm. To achieve further speedup, \cite{liu2023llms} transformed the high-overhead functions into cryptography-friendly approximations, and finetuned the model to maintain accuracy. SecFormer designs a combination of SMPC protocols for accurate and efficient computation of complex nonlinear functions, such as GeLU and LayerNorm \cite{luo2024secformer}. BOLT reduces the payload on both linear and non-linear operations through cryptographic optimizations, and eliminates insignificant words using ML techniques for enhanced efficiency \cite{pang2024bolt}. NEXUS introduces the first non-interactive protocol, allowing the client to complete the entire inference process with the server in a single round of communication \cite{zhang2024secure}. Despite their privacy advantages, encryption-based methods impose significant computational overhead and require substantial investment in hardware and algorithmic acceleration, making integration with existing black-box LLMs challenging.

\textit{Perturbation-based method.} Perturbation-based methods provide privacy guarantee by injecting calibrated noise into the input. Existing studies utilizing this method predominantly focus on privacy protection in fine-tuning \cite{kerrigan2020differentially,yu2021differentially,lidag} and prompt-tuning phases \cite{duan2023flocks,li2023privacypreserving}, while few studies investigate the privacy-preserving inference paradigm. A major challenge in perturbation-based private inference is to balance the utility and privacy trade-off. Recent studies have proposed Text2Text \cite{feyisetan2020privacy} and paraphraser-based approaches \cite{utpala2023locally,mattern2022limits} to privatize text with LDP. Split-N-Denoise (SnD) \cite{maisplit} deployed the token embedding layer at the client side, and introduced a user-side denoising model to correct the purturbed embedding output for downstream tasks. More recently, DYNTEXT proposes a semantic-aware dynamic text sanitization mechanism for privacy-preserving LLM inference, which adaptively perturbs sensitive tokens according to their semantic density to improve the privacy-utility trade-off \cite{zhang2025dyntext}. InferDPT further studies privacy-preserving inference for closed-box LLMs by combining a perturbation module with a local extraction module, such that a perturbed prompt is uploaded to the remote LLM and a lightweight local model reconstructs a higher-quality output from the perturbed generation \cite{tong2025inferdpt}. In a related but distinct setting, FedCoT proposes a federated chain-of-thought distillation framework that transfers knowledge from a server-side LLM to a client-side small language model using perturbed prompts and rationales to protect user privacy \cite{fan2025fedcot}. However, existing perturbation-based methods either significantly degrade utility or are incompatible with black-box LLMs due to proprietary and efficiency concerns.

In summary, existing approaches either significantly degrade model performance or lack compatibility with black-box LLMs. To address this, we propose a confusion-based strategy that uploads a mix of genuine and synthetic prompts while preserving privacy. Table \ref{tab:comppriv} summarizes the coarse-grained comparision between ConfusionPrompt and existing approaches.

\begin{table}[htp]
\caption{Coarse-grained comparison between privacy-preserving LLM inference frameworks.}
\label{tab:comppriv}
\begin{center}
\begin{sc}
\scalebox{0.93}{
\begin{tabular}{lccc}
\toprule
Method & Plug-and-Play & Impact on Utility \\
\midrule
HaS \cite{chen2023hide} & Yes & Medium \\
 Casper \cite{chong2024casper} & Yes & Medium \\
 \midrule
Iron \cite{hao2022iron} & No & Low  \\
 SecFormer \cite{luo2024secformer} & No & Low  \\
  BOLT \cite{pang2024bolt} & No & Low \\
  NEXUS \cite{zhang2024secure} & No & Low \\
 \midrule
Text2Text \cite{feyisetan2020privacy} & Yes & High  \\
  Paraphraser \cite{utpala2023locally,mattern2022limits} & Yes & High  \\
  DYNTEXT \cite{zhang2025dyntext} & Yes & Medium  \\
  InferDPT \cite{tong2025inferdpt} & Yes & Medium  \\
  SnD \cite{maisplit} & No & Medium \\
 \midrule
 \textbf{ConfusionPrompt} & \textbf{Yes} & \textbf{Low} \\
\bottomrule
\multicolumn{3}{l}{}\\[-5pt]
\multicolumn{3}
{@{\extracolsep{\fill}}p{22pc}@{\extracolsep{\fill}}}{\hspace*{9pt}\textnormal{ConfusionPrompt offers a plug-and-play solution with little impact on utility. It achieves superior privacy-utility trade-off compared to sanitization and perturbation methods, which often introduce semantic bias into the original prompt.}}
\end{tabular}
}
\end{sc}
\end{center}
\end{table}

\section{ConfusionPrompt: Design and Formulation}

In this section, we first illustrate the overview of our proposed ConfusionPrompt, which consists of three critical components: decomposer, generator, and recomposer.
Then, we develop a privacy model to formulate the requirements for privacy-preserving inference.

\subsection{Overall Design}

Denote $G_{s}: \mathcal{V}^* \rightarrow \mathcal{V}^*$ as the LLM deployed as cloud service such as ChatGPT \cite{ouyang2022training,openai2023gpt4,chen2025mark}, where $\mathcal{V}^*$ is the vocabulary space.
Instead of transmitting raw user prompt to the cloud, ConfusionPrompt introduces a privacy-preserving inference paradigm that significantly hinders third parties from inferring the original user prompt, even if they access the transmitted data.
Our framework consists of six key steps:

\emph{Step 1: decomposition of the original prompt.}
We first introduce a decomposer $G_d: \mathcal{V}^* \rightarrow \mathcal{V}^*$ that aims to decompose the user's original prompt $p$ to a sequence of genuine sub-prompts $\boldsymbol{p}=[p_1, p_2,...,p_{|\boldsymbol{p}|}]$, resulting in fewer private attributes (i.e., private information) in each sub-prompt.

\emph{Step 2: generation of pseudo prompts.}
A pseudo prompt generator $G_f: \mathcal{V}^* \rightarrow \mathcal{V}^*$ is introduced to generate a pseudo prompt given a genuine sub-prompt by replacing specified critical information.
By generating multiple pseudo prompts for each genuine sub-prompt and mixing them together as a prompt group, we are able to hide the genuine sub-prompts, thus increasing the difficulty of being inferred.
For the simplicity of privacy analysis, we design to enforce the pseudo prompt to be consistent with the genuine prompt in terms of syntactic structure.

\emph{Step 3: evaluation of privacy level.}
To ensure that the genuine prompt is safely hidden in the prompt group, we also need to design some criteria (e.g., semantic irrelevance) to evaluate the usability of generated pseudo-prompts.
Based on these criteria, we iteratively sample and filter the pseudo-prompts until the prompt group meets the privacy requirement, which will be introduced in details in Section \ref{sec:privacy_model}.

\emph{Step 4: communication with cloud server.}
The user transmits the prompt group to the server for LLM processing, and the server returns the corresponding response group. Throughout this process, the prompt group is designed to obscure the original user prompt, making it difficult for the server to infer the sensitives information, thereby safeguarding user's privacy.

\emph{Step 5: retrieval of interested response.}
Since the user knows which sub-prompts are genuine, they can effortlessly extract the corresponding sub-responses while discarding those from pseudo-prompts.

\emph{Step 6: recomposition of sub-responses.}
Here, we introduce a recomposer $G_r: \mathcal{V}^* \rightarrow \mathcal{V}^*$ that maps a sequence of sub-prompt-response pairs to a final response.

\begin{figure}[!htp]
    \centering
    \includegraphics[width=0.95\linewidth]{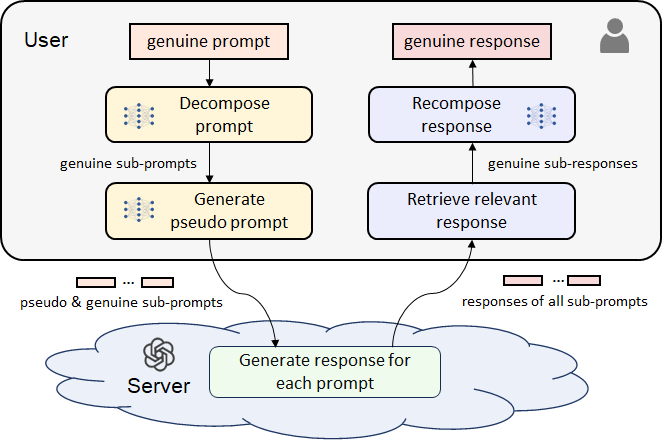}
    \caption{Overview of ConfusionPrompt. 
    }
    \label{fig:overview}
\end{figure}

\subsection{Privacy Model}
\label{sec:privacy_model}

\subsubsection{Rationale of Privacy Model}
In this section, we define a $(\lambda, \mu, \rho)$-privacy model to formulate the requirements that the group of pseudo-prompts should satisfy for privacy protection. This model provides guidance for evaluating privacy levels and training components (i.e., decomposer, generator, and recomposer). To explain the rationale of our privacy model, we follow \cite{pang2010embellishing} to quantify the privacy risk of queries exposed to the server.

Consider a set of prompts denoted as $\boldsymbol{P}=\{\boldsymbol{p}_1, \boldsymbol{p}_2, ..., \boldsymbol{p}_n\}$. For any $\boldsymbol{p}\in \boldsymbol{P}$, let $\pi(\boldsymbol{p})$ be the adversary's prior probability that $\boldsymbol{p}$ is the genuine prompt provided by the user. Then the adversary's posterior probability $\pi(\boldsymbol{p}|\boldsymbol{P})$ of the same event is:

\begin{equation}
    \pi(\boldsymbol{p}|\boldsymbol{P}) = \frac{\pi(\boldsymbol{p})}{\sum_{\boldsymbol{p}'\in \boldsymbol{P}} \pi(\boldsymbol{p}')}.
\end{equation}

Let $\boldsymbol{p}^0$ denote the genuine prompt provided by the user. Then the privacy risk associated with the set of prompts $\boldsymbol{P}$ revealed to the server can be formulated as:
\begin{equation}
\label{eq:risk}
    \mathrm{Risk}(\boldsymbol{P}) = \sum_{\boldsymbol{p}'\in \boldsymbol{P}} \pi(\boldsymbol{p}'|\boldsymbol{P}) \cdot 
    \mathrm{Sim} (\boldsymbol{p}', \boldsymbol{p}^0),
\end{equation}
where $\mathrm{Sim}$ is some measure of similarity between two prompts.

Equation \ref{eq:risk} suggests three possible methods to minimize the privacy risk. First, reducing $\mathrm{Sim} (\boldsymbol{p}', \boldsymbol{p}^0)$ for any $\boldsymbol{p}' \neq \boldsymbol{p}^0$ leads to a decrease in $\mathrm{Risk}(\boldsymbol{P})$. Therefore, generating deceptive prompts that are semantically distinct from the genuine prompt can effectively lower the privacy risk. Second, $\mathrm{Risk}(\boldsymbol{P})$ is reduced when $\pi(\boldsymbol{p}'|\boldsymbol{P})$ is high for prompts $\boldsymbol{p}'$ that do not resemble $\boldsymbol{p}^0$. This implies that fabricated prompts should be crafted to appear as realistic as possible. Finally, $\mathrm{Risk}(\boldsymbol{P})$ decreases if we increase the number of prompts $\boldsymbol{p}'$ with low $\mathrm{Sim} (\boldsymbol{p}', \boldsymbol{p}^0)$. This can be achieved by increasing the number of fake prompts with different semantic meanings.

Based on the observations, we develop the following criteria for the fabricated prompts: (i) The fabricated attributes in the pseudo prompt should be semantically irrelevant to the original attributes in genuine prompt, so that the genuine user information can be effectively obfuscated \cite{wu2015constructing}.
(ii) The genuine prompt should be obfuscated by sufficient number of pseudo-prompts, and the attacker can not easily identify the real prompt from the combination pattern even with background knowledge.
(iii) The pseudo-prompt should appear genuine rather than obviously fabricated. Thus it is crucial to maintain the fluency and reasonability of the pseudo-prompts.

\subsubsection{Construction of Privacy Model}
In the following, we first give the definitions of private attributes and their semantic relevance.
\begin{definition}[Private Attributes]
\label{def:private}
Denote $\mathcal{U}$ as the attribute space, and $\mathcal{P}$ as the prompt space.
Given a sub-prompt list $\boldsymbol{p} \in \mathcal{P}$, its private attributes are denoted as $U(\boldsymbol{p}) = \{u_1, u_2, ..., u_m\}$ for $m$ attributes.
\end{definition}
\begin{remark}
For conciseness, the prompt in Definition \ref{def:private} could be a full prompt $p$, or a sequence of decomposed prompts with the same intention $\boldsymbol{p}=[p_1, p_2,...,p_{|\boldsymbol{p}|}]$.
\end{remark}
The private attributes encompass a range of elements, including verbs, adjectives, and nouns, which extend beyond personally identifiable information (PIIs). Next, we define the semantic similarity between two private attributes.

\begin{definition}[Attribute-attribute Similarity]
Given two attributes $u_1$, $u_2\in \mathcal{U}$, an attribute similarity function measures the semantic similarity between $u_1$ and $u_2$: $\text{Sim}(u_1, u_2): \mathcal{U} \times \mathcal{U} \rightarrow \mathbb{R}$.
\end{definition}

For a pair of genuine and pseudo-prompts, we define the correspondent attributes as follows:
\begin{definition}[Correspondent Attributes]
Given two prompts $\boldsymbol{p}_1$, $\boldsymbol{p}_2\in \mathcal{P}$, the correspondent attribute of $u_i\in U(\boldsymbol{p}_1)$ in $\boldsymbol{p}_2$, is defined as the attribute at the same syntactic position of $\boldsymbol{p}_2$, denoted by $\text{Corr}(u_i, \boldsymbol{p}_1, \boldsymbol{p}_2)$.
\end{definition}
\begin{remark}
The definition assumes that: (1) $\boldsymbol{p}_1$ and $\boldsymbol{p}_2$ have the same syntactic structure; (2) if $u_i$ occurs in $\boldsymbol{p}_1$ multiple times, the correspondent locations in $\boldsymbol{p}_2$ return the same attribute.
\end{remark}
For example, given the original question "What are the responsibilities of software engineers?" and private attribute "software engineer", the correspondent attribute in the fake question "What are the responsibilities of school teachers?" would be "school teacher". Based on the correspondent attributes and attribute-attribute similarity, we can define the semantic similarity between two prompts below:

\begin{definition}[Prompt-prompt Similarity]
\label{def:promptsim}
Given two prompts $\boldsymbol{p}_1$, $\boldsymbol{p}_2\in \mathcal{P}$, their similarity can be defined by the similarity between each pair of correspondence attributes: 
\begin{equation}
    \mathrm{Sim}(\boldsymbol{p}_1, \boldsymbol{p}_2) = \max_{u_i \in U(\boldsymbol{p}_1)}  \text{Sim}\left(u_i, \text{Corr}(u_i, \boldsymbol{p}_1, \boldsymbol{p}_2)\right)
\end{equation}
\end{definition}

Next, we define the significance of an attribute, which is tied to the likelihood of a curious server being able to identify the genuine attribute from a group of genuine and pseudo-prompts.
\begin{definition}[Significance of Single Attribute]
Denote $\boldsymbol{P}=\{\boldsymbol{p}_1, \boldsymbol{p}_2, ..., \boldsymbol{p}_n\}$ as a group of prompts. The significance of an attribute $u\in U(\boldsymbol{p}_i)$ related to a group of prompts $\boldsymbol{P}$ is defined as:
\begin{equation}
    \mathrm{Sig}(u, \boldsymbol{P}) = \frac{\sum_{j=1}^n H(u, \boldsymbol{p}_i, \boldsymbol{p}_j)}{n},
\end{equation}
where:
\begin{equation}
H(u, \boldsymbol{p}_i, \boldsymbol{p}_j) =\left\{
\begin{array}{ll}
1 & \text{Corr}\left(u, \boldsymbol{p}_i, \boldsymbol{p}_j\right)=u\\
0 & \text{otherwise}
\end{array}.
\right.
\end{equation}
\end{definition}
In our setting, the group of prompts consists of one genuine prompt along with a collection of pseudo-prompts.
The significance of a true attribute can be considered as the proportion of its occurrence within the group of prompts.
In the following, we provide the definition of significance for multiple attributes.

\begin{definition}[Significance of Attribute Set]
\label{def:sigset}
Suppose $\boldsymbol{p}_0$ denotes the genuine prompt. The significance of attribute set $U(\boldsymbol{p}_i)$ related to a group of prompts $\boldsymbol{P}=\{\boldsymbol{p}_0, \boldsymbol{p}_1, ..., \boldsymbol{p}_n\}$ is defined as:

\begin{equation}
\begin{aligned}
    &\mathrm{Sig}(U(\boldsymbol{p}_0), \boldsymbol{P}) \\
    =&
    \max_{u_k}  \max_h 
    \max_{\mathcal{V}} 
    \frac{\sum_{j=1}^n \bar{H}(\mathcal{V}, p_{0h}, p_{jh})}{\sum_{j=1}^n \bar{H}(\mathcal{V}\backslash u_k, p_{0h}, p_{jh})} \\
    &{\rm s.t.} u_k \in \mathcal{V}\subseteq U(p_{0h}),
\end{aligned}
\end{equation}
where:
\begin{equation}
\bar{H}(\mathcal{V}, p_{0h}, p_{jh}) =\left\{
\begin{array}{ll}
1 & \mathrm{Corr}\left(u,p_{0h}, p_{jh}\right)=u\ \forall u \in \mathcal{V} \\
& \mathrm{or}\ \mathcal{V}=\emptyset\\
0 & \mathrm{otherwise}
\end{array},
\right.
\end{equation}
\end{definition}
$p_{ih}$ is the $h$-th decomposed sub-prompt in the $i$-th prompt.
In case that the attacker have prior knowledge of some private attributes, we introduce a notation $\mathcal{V}$ as a subset of private attributes in $p_{0h}$.
Then we consider the proportion of a given attribute conditioned on any possible set of private attributes to bound the possibility of correctly identifying the target attribute with some background knowledge.
Now, we provide the definition of genuineness for the third criterion:
\begin{definition}[Genuineness]
\label{def:genuine}
Denote $D: \mathcal{P} \rightarrow \mathbb{R}$ as a function to discriminate between fabricated and genuine prompts. A larger value returned by $D$ indicates a higher likelihood of a prompt being genuine. The genuineness of a prompt $p\in \mathcal{P}$ can be defined as $\text{Genu}(p) = D(p)$.
\end{definition}

Based on Definition \ref{def:promptsim}, \ref{def:sigset}, and \ref{def:genuine}, we can formulate our proposed $(\lambda, \mu, \rho)$-privacy model as:
\begin{definition}[User Privacy]
Let $\boldsymbol{P}=\{\boldsymbol{p}_0, \boldsymbol{p}_1, ..., \boldsymbol{p}_n\}$ be a group of prompts, where $\boldsymbol{p}_0$ is the genuine user prompt and the remaining ones are pseudo-prompts. If $\boldsymbol{P}$ satisfies the following requirements, then it is deemed that it can ensure $(\lambda, \mu, \rho)$-privacy of user prompt $\boldsymbol{p}_0$ with respect to discriminator $D$:
\begin{itemize}
    \item Each pseudo prompt should be semantically irrelavant to user prompt $\boldsymbol{p}_0$, i.e., $\forall i\in [1,n]$, $\text{Sim}(\boldsymbol{p}_0, \boldsymbol{p}_i) \leq \lambda$.
    \item The users' sensitive attributes should be obfuscated by sufficient pseudo-prompts, i.e., $\text{Sig}(U(\boldsymbol{p}_0), \boldsymbol{P})\leq \mu$. 
    \item Each pseudo-prompt should not be classified as fabricated prompt by discriminator $D$, i.e., $\forall i\in [1,n]$, $\text{Genu}(p_i) = D(p_i) \geq \rho$.
\end{itemize}
\end{definition}
In \ref{app:attacksig}, we provide the relationship between our privacy model and the upper bound on inference attacks.

\subsubsection{Implication and Selection of Privacy Parameters}
In this section, we provide a explanation for the three parameters in our privacy model:
\begin{itemize}
    \item Similarity parameter $\lambda$ controls the similarity between private attributes in genuine and pseudo-prompts. A lower $\lambda$ results in more dissimilar pseudo-prompts, thereby enhancing privacy protection. In our empirical analysis, we measure cosine similarity in the embedding space of private attributes. As shown in Figure \ref{fig:attrattack}, reducing $\lambda$ decreases the success rate of attribute inference attacks, with attack accuracy remaining sufficiently low for $\lambda<0.8$ under appropriate choices of the other two parameters (i.e., $\mu>10$ and $\rho=4$).
    \item Significance parameter $\mu$ provides the bound on attack success rate of correctly identifying the target attribute. A lower $\mu$ necessitates generating a larger number of pseudo-prompts to obscure sensitive information, while providing higher level of privacy protection. The choice of $\mu$ depends on the user's privacy requirements and the attribute space conditioned on the non-private information.
    \item Genuineness parameter $\rho$ quantifies the degree of linguistic realism exhibited by individual synthetic prompts. In empirical analysis, we define the genuineness of a prompt on a four-point ordinal scale: 1 (incomprehensible), 2 (low quality), 3 (moderate fluency), and 4 (perfect fluency). We recommend the user to configure $\rho=4$, i.e., requiring each pseudo-prompt to satisfy perfect fluency, so that the attacker can not easily detect fabricated prompts.
\end{itemize}

\section{ConfusionPrompt: User-Side Models}

Following the overall design and the requirements in privacy model, this section introduces how to enable ConfusionPrompt through the design of user-side models, including decomposer, generator, and recomposer.

\subsection{Decomposer}

A decomposer is designed to decompose an original user prompt into several sub-prompts, such that the required number of pseudo-prompts to ensure the same level of privacy preservation can be significantly reduced.
In the following, we theoretically demonstrate the benefit of a local decompose module in terms of complexity measured by the number of required pseudo-prompts.

\subsubsection{Complexity Analysis}

We first show the complexity for single-paragraph prompt, that is, the prompt is not decomposed.

\begin{theorem}[Complexity for Single Paragraph]
\label{theo:complexsingle}
Let $\boldsymbol{P}=\{\boldsymbol{p}_0, \boldsymbol{p}_1, ..., \boldsymbol{p}_n\}$ be a group of prompts, where $\boldsymbol{p}_0$ is the genuine user prompt and the remaining ones are dummy prompts. Suppose each prompt in the group represents a single paragraph $|\boldsymbol{p}_i|=1$, $\forall i \in [1,n]$, meaning that each contains a single query. To achieve $(\lambda, \mu, \rho)$-privacy, it requires at least $n$ prompts:
\begin{equation}
    n\geq \left(\frac{1}{\mu}\right)^{|U(\boldsymbol{p}_0)|},
\end{equation}
where $|U(\boldsymbol{p}_0)|$ represents the number of private attributes in prompt $\boldsymbol{p}_0$.
\end{theorem}

Accordingly, we provide the following complexity analysis for decomposed prompts, which are a key component in our ConfusionPrompt.

\begin{figure}[htbp]
    \centering
    \includegraphics[width=0.98\columnwidth]{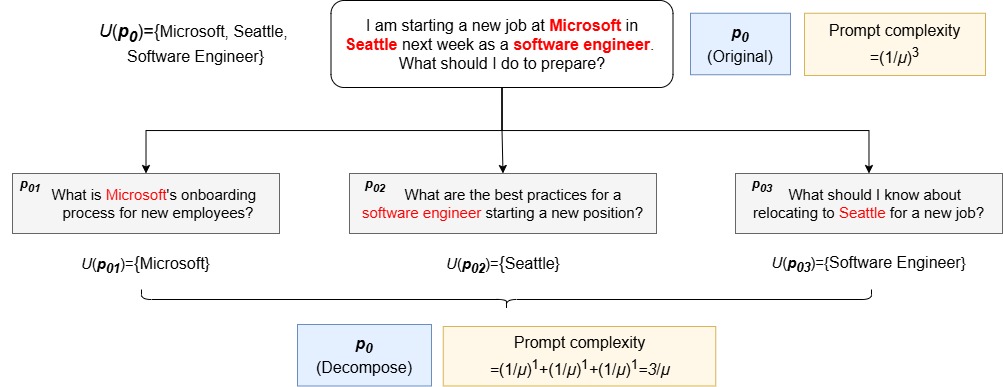}
    \caption{Real world example illustrating Theorem \ref{theo:complexsingle} and \ref{theo:complexdecomp}. The complexity is reduced from $(1/\mu)^3$ to $3/\mu$.}
    \label{fig:decompexamreal}
\end{figure}

\begin{theorem}[Complexity for Decomposed Prompt]
\label{theo:complexdecomp}
Let $\boldsymbol{P}=\{\boldsymbol{p}_0, \boldsymbol{p}_1, ..., \boldsymbol{p}_n\}$ be a group of prompts, where $\boldsymbol{p}_0$ is the user prompt and the remaining ones are dummy prompts. Suppose each prompt in the group represents a sequence of decomposed prompts with the same intention $\boldsymbol{p}_i=[p_{i1}, p_{i2}, ..., p_{il}]$, $\forall i \in [1,n]$. To achieve $(\lambda, \mu, \rho)$-privacy, it requires at least $n$ prompts:
\begin{equation}
\label{eq:ndecomp}
    n\geq \sum_{j=1}^l \left(\frac{1}{\mu}\right)^{|U(p_{0j})|},
\end{equation}
where $|U(p_{0j})|$ is the number of private attributes in the $j^{th}$ decomposed query, and $l$ denotes the number of decomposed queries in a prompt.

Specifically, if $|U(\boldsymbol{p}_0)|=l$ and $|U(p_{0j})|=1$, $\forall j\in [1,l]$, meaning that each decomposed query contains a single attribute, then
\begin{equation}
    n\geq \frac{|U(\boldsymbol{p}_0)|}{\mu}
\end{equation}
\end{theorem}
Theorem \ref{theo:complexsingle} and \ref{theo:complexdecomp} compare the complexities between single prompt and decomposed sub-prompt, where the proof is defered to \ref{app:complexproof}. In the ideal scenario, the complexity can be reduced to $\mathcal{O}\left(|U(\boldsymbol{p}_0)|/\mu\right)$ after decomposition. An intuitive explanation is given in Figure \ref{fig:decompexamreal}, where a raw query containing three private attributes is decomposed into three sub-questions, each including a single attribute.

\subsubsection{Requirements of Ideal Decomposer}
Now, we can articulate the criteria for an ideal decomposer: 
\begin{itemize}
    \item A fundamental requirement for the decomposition process is to follow the collectively exhaustive principle \cite{rasiel1999mckinsey}. In other words, the collection of responses to decomposed sub-prompts should recover the whole response. 
    \item An ideal decomposer would optimize the complexity, i.e., the required number of pseudo-prompts, in two aspects: (i) each sub-prompt contains as few attributes as possible; (ii) each attribute appears in as few prompts as possible.
\end{itemize}

\subsubsection{Model Training}
\label{sec:traindecomp}
We design a two-stage training approach to optimize the above objectives. In the first stage, we finetune a pretrained language model (LM) on demonstration data for decomposition task, where the LM is trained to generate decomposition given a full prompt. In the second stage, the LM is prompted to output multiple decompositions for each input. From the generations, a subset of preferred decompositions is selected to further finetune the model. The procedure to select preferred examples is as follows:
\begin{itemize}
    \item Evaluate whether each decomposition adheres to the collectively exhaustive principle, discarding any that do not meet this standard.
    \item Among the qualified decompositions, assess the complexity of each using a predefined parameter $\mu$, and select the top $k$ examples for each prompt with the lowest complexity.
\end{itemize}

\subsection{Generator}
Generator takes an genuine prompt $p$ and its private attributes $U(p)$ as input and produces a pseudo prompt $p'$ with replaced attributes.
The generator will be run for multiple times to obtain a series of pseudo-prompts for each genuine prompt.

\subsubsection{Requirements}
The requirements for a pseudo-prompt suggest the following criteria for an ideal generator:  
\begin{itemize}
    \item The generator should produce pseudo-prompt with fake information that is semantically irrelevant to the correspondent attributes in genuine prompt.
    \item The pseudo-prompt should not be classified as fabricated prompt by a strong discriminator $D$.
\end{itemize}

\subsubsection{Model Training}
The generator is trained using a similar procedure described in Section \ref{sec:traindecomp}. In the first stage, we finetune a pretrained LM using demonstration data for replacement task, where the LM is trained to generate new prompt with replaced attributes. In the second stage, we collect the preferred generations considering both semantic relevance and genuineness, with criteria given as follows:
\begin{itemize}
    \item Compute $s$, the similarity between the fabricated and original prompt.
    \item Compute $f$, the genuineness score for the pseudo-prompt.
    \item Calculate the overall score $\beta f-s$, where $\beta$ is the hyperparameter controlling the weight of the genuineness score.
    \item Select the top $k$ examples for each prompt with the highest overall score.
\end{itemize}

\subsection{Recomposer}

The local re-composition LM combines the retrieved sub-prompts and sub-responses to produce the final whole response.
We train a local recomposer on a collection of demonstration data with supervise learning.
Specifically, given a complete prompt $p$ and a set of sub-prompt-response pairs $\left((p_1, r_1), ...(p_k, r_k)\right)$, the recomposer $G_r$ is trained to maximize $\log p_{G_r}\left(r|p, (p_1, r_1), ...(p_k, r_k)\right)$ where $r$ is the final response.

\subsection{Algorithm}
Algorithm \ref{alg:framework} outlines the procedure for our private inference framework ConfusionPrompt.
The client starts with decomposing the full prompt into sub-prompts and then generates a collection of qualified fake prompts that meet the privacy requirement.
It can be derived immediately from the algorithm that our protocol meets $(\lambda, \mu, \rho)$-privacy.

\begin{algorithm}[htp]
   \caption{ConfusionPrompt}
   \label{alg:framework}
\begin{algorithmic}[1]
   \Statex \textbf{Input:} Original prompt $p$, privacy parameters $\lambda$, $\mu$, and $\rho$.
   \Statex \textbf{Output:} Final response $r$.
   \State User decomposes the original prompt into sub-tasks $\boldsymbol{p}_0 = G_d(p)$.
   \For{$p_{0h}\in \boldsymbol{p}_0$} 
   \State User keeps sampling from the generator until there are $n$ distinct pseudo prompts with $n$ given by \ref{eq:ndecomp}, where each one $1\leq i \leq n$ has prompt similarity $\mathrm{Sim}(p_{0h}, p_{ih})\leq \lambda$ and genuineness score $\mathrm{Genu}(p_{ih})\geq \rho$.
   \EndFor
   \State User sends the group of genuine and pseudo prompts $\boldsymbol{P}=\{\boldsymbol{p}_0,\boldsymbol{p}_1,...,\boldsymbol{p}_n\}$ to the server.
   \State Server returns a collection of responses $(\boldsymbol{r}_0,...,\boldsymbol{r}_n)$.
   \State User retrieves the response corresponding to the genuine prompt $\boldsymbol{r}_0$.
   \State User obtains the final response $r$ using the local recomposition model.
\end{algorithmic}
\end{algorithm}

In the following, we analyze the computational complexity for the user and server.
\begin{itemize}
    \item User computation: For $\boldsymbol{p}_0=[p_{01}, p_{02}, ..., p_{0l}]$, the computation complexity can be broken as: (i) decompose the original prompt $O(l)$; (ii) generate pseudo-prompts $O\left(1/\alpha\sum_{j=1}^l \left(1/\mu\right)^{|U(p_{0j})|}\right)$, where $\alpha$ denotes the probability that a generated prompt satisfies the privacy requirement. In the ideal decomposition scenario, this reduces to $O\left(|U(\boldsymbol{p}_0)|/(\alpha\mu)\right)$. (iii) recompose the responses $O(l)$. The primary computational overhead arises from pseudo-prompt generation.
    \item Server computation: The server computation complexity is $O\left(\sum_{j=1}^l \left(1/\mu\right)^{|U(p_{0j})|}\right)$, which, in the ideal decomposition scenario, reduces to $O\left(|U(\boldsymbol{p}_0)|/\mu\right)$. This directly corresponds to the user query cost.
\end{itemize}

\section{Experiment}
\subsection{Experiment Setup}
We evaluate our framework on three datasets: StrategyQA \cite{geva2021did}, MuSiQue \cite{trivedi2022musique}, and MMLU \cite{hendryckstest2021}. These datasets encompass questions from a wide range of areas, such as business, medicine, and religion (see \ref{app:cat}).
We integrate ConfusionPrompt with different online LLMs including GPT-3.5-Turbo, GPT-4-Turbo, and GPT-4o \cite{openai2023gpt4}. We consider the privacy parameters $\lambda \in [0.5, 0.8]$, $1/\mu \in [5, 
50]$, and $\rho \in [1, 4]$ unless specified. Refer to \ref{app:localtrain} for the training of decomposer, generator, and recomposer.

We extract the private attributes of the three datasets through the steps below (see \ref{app:attrconst} for details):
\begin{itemize}
    \item Step 1: sample construction. We adopt a semi-automated method to extract private attributes from each sample. Specifically, we extract the entities using a combination of two NER methods: Spacy \cite{srinivasa2018natural} and Flair \cite{akbik2019flair}. Then we manually correct and supplement the private attributes for each query.
    \item Step 2: model finetuning. An LLM is finetuned using the private attribute samples so that it outputs private attributes given a query.
    \item Step 3: attribute extraction. The remaining queries are fed into the finetuned LLM to generate private attributes.
\end{itemize}

\subsection{Empirical Privacy Evaluation}
We simulate privacy attacks to investigate the privacy protection level under various combinations of privacy parameters. Existing privacy attacks can be categorized as: (i) membership inference attack that identifies whether a record is included in a model's training dataset \cite{shokri2017membership, hu2022membership}, (ii) attribute inference attack that infers specific sensitive attributes \cite{bhagoji2019analyzing, ateniese2015hacking}, and (iii) reconstruction attack that reconstructs the original data \cite{zhu2019deep, he2019model}. 

The first attack is more associated with training time rather than inference time. Therefore, we focus on the latter two attacks, which are adapted to our ConfusionPrompt framework as follows:

\textit{Prompt Identification Attack:} an attack that identifies the true query from a group of fake and true prompts. It is an adaptation of the reconstruction attack to our framework, where the attacker's view is a set of prompts. Denote $\mathbf{q}=[\![q_1||q_2||...||q_n]\!]$ as the concatenated group of queries, and $k$ as the index for the true query.  We finetune BART-large and Qwen3.5-4B classification models, denoted as $G_{\text{PIA}}$, to predict the index of the true query $\hat{k}=G_{\text{PIA}}(\mathbf{q})$. 

\textit{Attribute Inference Attack:} an attack that infers the sensitive features of records from either a group of prompts, or differentially privatized inputs. We rely on the twitter text dataset \cite{vashisth2020gender} to predict user gender based on the review.

Noted that prompt identification attack is specifically designed for our framework, and attribute inference attack can be applied to both ConfusionPrompt and DP-based methods.

\subsection{Experiment Results}

\subsubsection{Privacy Experiments}
\label{sec:privacyexp}
\begin{figure}[htp]
\begin{center}
\includegraphics[width=1\linewidth]{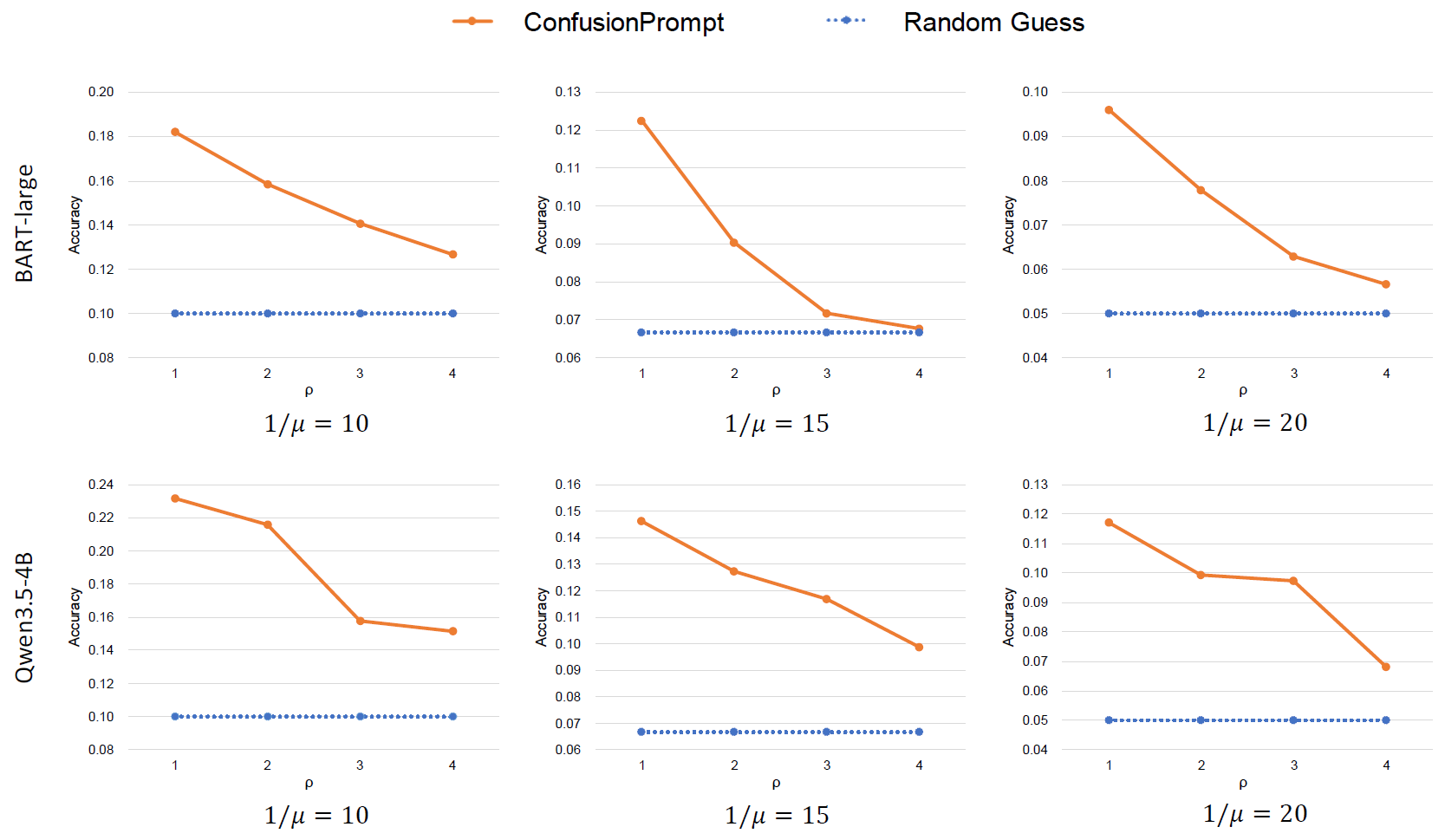}
\caption{Prompt identification attack accuracy under various combinations of significance parameter $\mu$ and genuineness parameter $\rho$ for two attack models. Similarity parameter $\lambda$ is fixed at 0.5.}
\label{fig:piattack}
\end{center}
\end{figure}

Figure \ref{fig:piattack} visualizes the attack accuracy for prompt identification attack under various combinations of significance $\mu$ and genuineness $\rho$. It can be observed that: (a) Decreasing significance $\mu$ consistently reduces the attack accuracy. (b) As the genuineness threshold $\rho$ increases from 1 to 4, the attack accuracy approaches that of random guessing. The result reveals that it becomes harder for the attacker to distinguish the true query as the fake prompts become more realistic. In Figure \ref{fig:piattackapp}, we investigate the impact of similarity threshold $\lambda$ on the attack success rate, and find no clear relationship between the two (see \ref{app:attack}). Such privacy parameter could be more related to the attribute inference attack discussed later.

\begin{figure}[htp]
\begin{center}
 \includegraphics[width=0.95\columnwidth]{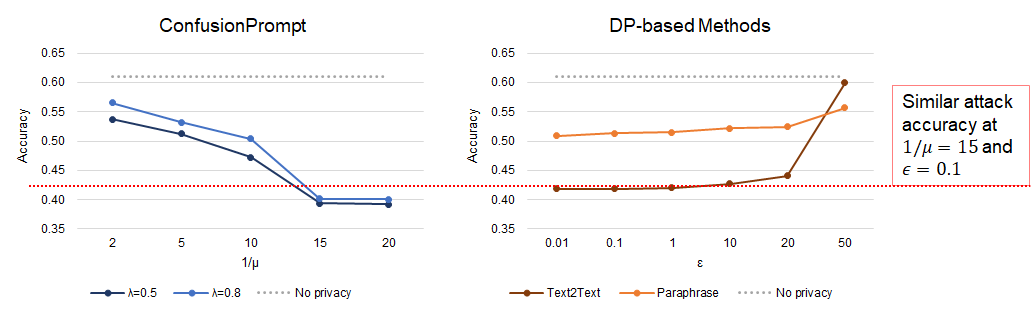}
\caption{Attribute inference attack accuracy for ConfusionPrompt and LDP-based methods. Genuineness requirement $\rho$ is fixed at 4.}
\label{fig:attrattack}
\end{center}
\end{figure}

Figure \ref{fig:attrattack} presents the attack accuracies of attribute inference attack for both ConfusionPrompt and LDP-based methods, from which we can make the following observations: (1) Paraphrase generally leads to higher attack accuracies compared to Text2text. This can be attributed to the fact that the semantic meanings of certain key words remain unchanged during the paraphrasing process. (2) The attack accuracies decrease to a plateau when $1/\mu \geq 15$ or $\epsilon \leq 1$. This suggests that beyond these thresholds, further increasing the privacy protection metrics does not significantly reduce the attack accuracy. (3) The attack accuracy for ConfusionPrompt with $1/\mu \geq 15$ is similar to that of Text2Text with $\epsilon \leq 1$, indicating certain alignment for the privacy protection between LDP-based methods and our framework. (4) Decreasing the similarity threshold $\lambda$ helps to reduce the attribute inference attack accuracies.

\subsubsection{Utility Evaluation}
\label{sec:utility}
For StrategyQA, we report the accuracy scores (ACC) and area under the roc curve (AUC) to assess the the classification task. For MuSiQue, we report the F1 score, ROUGE-L, and exact matching (EM) to evaluate the question answering task \cite{yang2018hotpotqa}. For MMLU, we report the ACC and F1 score to evaluate the multiple-choice task. We benchmark our framework with $\mu=15$, $\rho=4$, and $\lambda=0.5$ against three categories of baseline methods: 
\begin{itemize}
    \item Local LLM. As ConfusionPrompt requires users to maintain multiple LMs for collaborative inference, it is essential to compare the performance of: (i) ConfusionPrompt utilizing online LLMs, and (ii) local inference using LLMs of comparable size to the LMs maintained on the user side. We thus evaluate the utility on two open-source LLMs: Llama2 \cite{touvron2023llama} model with 7B parameters and Vicuna \cite{chiang2023vicuna} model with 13B parameters.
    \item LDP approaches. We consider two LDP approaches integrable with black-box LLMs: (i) Text-to-text privatization (Text2Text) \cite{feyisetan2020privacy}, where the query is privatized with LDP by replacing each word with the perturbed token embeddings, and (ii) Paraphraser \cite{utpala2023locally, mattern2022limits} that paraphrases the prompt with LDP mechanism using a language model. The queries are privatized with LDP privacy budget $\epsilon=10$ before transmission to the server.
    \item Direct Query that queries online GPTs with no privacy protection. The complete prompt is transmitted to the service provider without any confusion strategy.
\end{itemize}

\begin{table*}[htp]
\caption{Utility comparisons between our proposed ConfusionPrompt and baselines.}
\label{tab:utility}
\begin{center}
\begin{sc}
\begin{tabular}{lc|cc|ccc|cc}
\toprule
\multirow{2}{*}{Method} & \multirow{2}{*}{Private} & \multicolumn{2}{c|}{StrategyQA} & \multicolumn{3}{c|}{MuSiQue} & \multicolumn{2}{c}{MMLU} \\
& & ACC & AUC & F1 & roughL & EM & ACC & F1 \\
\midrule
Llama2-7B & Yes& 0.60\tiny$\pm$0.02 & 0.58\tiny$\pm$0.01 & 0.48\tiny$\pm$0.02 & 0.48\tiny$\pm$0.02 & 0.32\tiny$\pm$0.01 & 0.56\tiny$\pm$0.03 & 0.54\tiny$\pm$0.02 \\
Vicuna-13B & Yes&0.65\tiny$\pm$0.00 & 0.63\tiny$\pm$0.01 & 0.42\tiny$\pm$0.01 & 0.42\tiny$\pm$0.01 & 0.23\tiny$\pm$0.02 & 0.60\tiny$\pm$0.00 & 0.60\tiny$\pm$0.01 \\
\midrule
Text2Text (GPT-3.5-Turbo) & Yes&0.53\tiny$\pm$0.04 & 0.51\tiny$\pm$0.05 & 0.02\tiny$\pm$0.00 & 0.02\tiny$\pm$0.01 & 0.01\tiny$\pm$0.01 & 0.37\tiny$\pm$0.02 & 0.36\tiny$\pm$0.02 \\
Text2Text (GPT-4-Turbo) & Yes&0.54\tiny$\pm$0.03 & 0.51\tiny$\pm$0.03 & 0.03\tiny$\pm$0.04  & 0.03\tiny$\pm$0.02 & 0.01\tiny$\pm$0.02 & 0.45\tiny$\pm$0.03 & 0.45\tiny$\pm$0.02\\
Text2Text (GPT-4o) & Yes&0.53\tiny$\pm$0.01 & 0.50\tiny$\pm$0.01 & 0.02\tiny$\pm$0.01 & 0.02\tiny$\pm$0.01 & 0.01\tiny$\pm$0.00 & 0.48\tiny$\pm$0.02 & 0.47\tiny$\pm$0.01 \\
\midrule
Paraphraser (GPT-3.5-Turbo) & Yes&0.49\tiny$\pm$0.02 & 0.48\tiny$\pm$0.02 & 0.08\tiny$\pm$0.00 & 0.08\tiny$\pm$0.00 & 0.03\tiny$\pm$0.01 & 0.47\tiny$\pm$0.08 & 0.46\tiny$\pm$0.10 \\
Paraphraser (GPT-4-Turbo) & Yes&0.56\tiny$\pm$0.09 & 0.53\tiny$\pm$0.06 & 0.06\tiny$\pm$0.01 & 0.06\tiny$\pm$0.01 & 0.03\tiny$\pm$0.00 & 0.58\tiny$\pm$0.05 & 0.57\tiny$\pm$0.06 \\
Paraphraser (GPT-4o) & Yes&0.53\tiny$\pm$0.11 & 0.51\tiny$\pm$0.08 & 0.07\tiny$\pm$0.01 & 0.07\tiny$\pm$0.01 & 0.04\tiny$\pm$0.00 & 0.59\tiny$\pm$0.02 & 0.58\tiny$\pm$0.02  \\
\midrule
Direct Query (GPT-3.5-Turbo) & No&0.71\tiny$\pm$0.01 & 0.74\tiny$\pm$0.01 & 0.60\tiny$\pm$0.00 & 0.61\tiny$\pm$0.00 & 0.44\tiny$\pm$0.01 & 0.74\tiny$\pm$0.00 & 0.74\tiny$\pm$0.01 \\
Direct Query (GPT-4-Turbo) & No&0.80\tiny$\pm$0.02 & 0.80\tiny$\pm$0.02 & 0.66\tiny$\pm$0.00 & 0.66\tiny$\pm$0.01 & 0.50\tiny$\pm$0.00 & 0.89\tiny$\pm$0.02 & 0.71\tiny$\pm$0.03   \\
Direct Query (GPT-4o) & No&0.79\tiny$\pm$0.00 & 0.78\tiny$\pm$0.01 &0.72\tiny$\pm$0.00 & 0.72\tiny$\pm$0.00 & 0.56\tiny$\pm$0.00 & 0.92\tiny$\pm$0.00 & 0.92\tiny$\pm$0.00 \\
\midrule
\textbf{ConfusionPrompt (GPT-3.5-Turbo)} & Yes&0.72\tiny$\pm$0.02 & 0.73\tiny$\pm$0.02 & 0.61\tiny$\pm$0.01 & 0.61\tiny$\pm$0.01 & 0.45\tiny$\pm$0.00 & 0.71\tiny$\pm$0.01 & 0.71\tiny$\pm$0.01 \\
\textbf{ConfusionPrompt (GPT-4-Turbo)} & Yes&0.74\tiny$\pm$0.01 & 0.74\tiny$\pm$0.02 & 0.63\tiny$\pm$0.02 & 0.63\tiny$\pm$0.03 & 0.50\tiny$\pm$0.02 & 0.83\tiny$\pm$0.02 & 0.82\tiny$\pm$0.02\\
\textbf{ConfusionPrompt (GPT-4o)} & Yes&0.73\tiny$\pm$0.03 & 0.74\tiny$\pm$0.03 & 0.68\tiny$\pm$0.02 & 0.68\tiny$\pm$0.01 & 0.54\tiny$\pm$0.01 & 0.89\tiny$\pm$0.01 & 0.90\tiny$\pm$0.02 \\
\bottomrule
\multicolumn{9}{l}{}\\[-5pt]
\multicolumn{9}
{@{\extracolsep{\fill}}p{42pc}@{\extracolsep{\fill}}}{\hspace*{9pt}\textnormal{The privacy parameters for ConfusionPrompt are fixed at $\mu=15$, $\rho=4$, and $\lambda=0.5$. The privacy budgets for LDP-based methods are set to $\epsilon=10$ and $\delta=10^{-4}$. Each value denotes the mean $\pm$ standard deviation under four rounds of experiments.}}
\end{tabular}
\end{sc}
\end{center}
\end{table*}

The results are reported in Table \ref{tab:utility}.
From the results, we can see that
(1) online proprietary models outperform open-source models by over 11\%, 24\%, and 18\% on StrategyQA, MuSiQue, and MMLU, respectively, indicating the strong motivation for users to leverage online service while preserving privacy.
(2) ConfusionPrompt based on various online models consistently and significantly outperforms LDP-based methods by an average of 42\%, 22$\times$, and 69\% on StrategyQA, MuSiQue, and MMLU, respectively, demonstrating the effectiveness of our proposed approach. Notably, our framework possesses a distinct advantage over perturbation-based methods: increasing the privacy protection level does not adversely affect performance, since the true queries are always provided to the server. 

\subsubsection{User Query Cost}
Theorem \ref{theo:complexsingle} and \ref{theo:complexdecomp} show that the decomposition module could reduce the number of pseudo-prompts required to be sent to the server. In this section, we empirically demonstrate practical benefits by analyzing the monetary cost of our ConfusionPrompt framework. In particular, we compute the monetary ratio, the ratio of the monetary cost for using ConfusionPrompt compared to the direct query method, both before and after the application of the decomposition module.

\begin{figure}[htp]
\begin{center}
\includegraphics[width=1\linewidth]{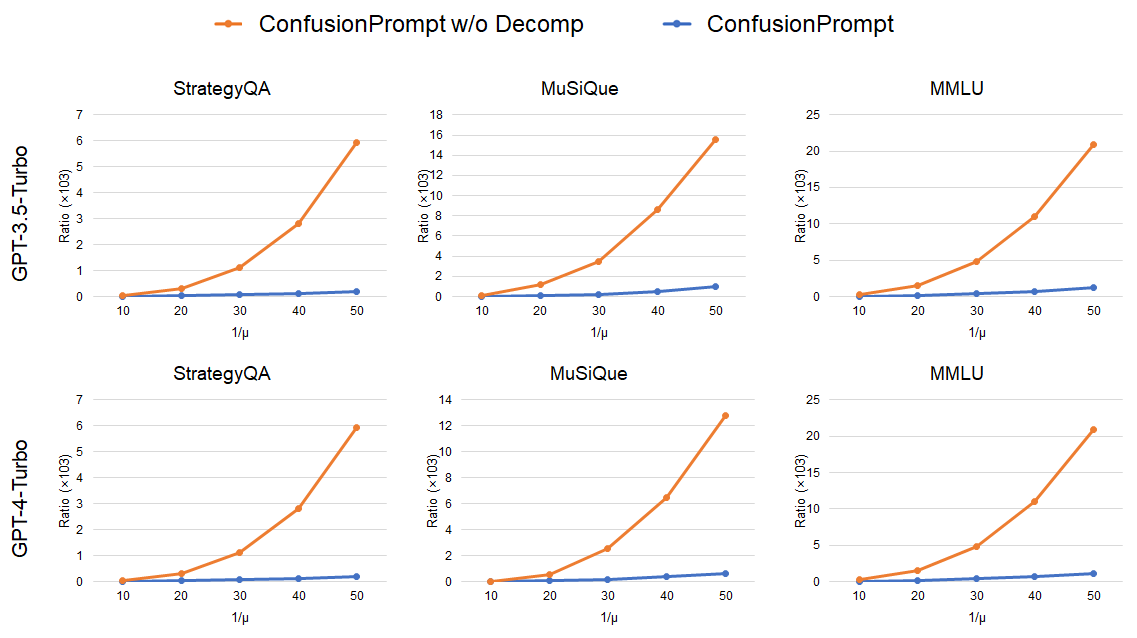}
\caption{Monetary ratio of strategyQA, MuSiQue, and MMLU datasets for ConfusionPrompt and the framework without decomposition (ConfusionPrompt w/o Decomp) under various significance $\mu$. Decompostion in ConfusionPrompt substantially reduces the monetary cost, indicating its efficiency.}
\label{fig:decompcomplexity}
\end{center}
\end{figure}

Figure \ref{fig:decompcomplexity} presents the query cost benefits achieved through our decomposition module.
It can be observed that the complexity gains become increasingly pronounced as the level of privacy protection increases. This can be attributed to the fact that the complexity experiences polynomial growth in $1/\mu$ before decomposition, whereas it exhibits approximate linearity dependency on $1/\mu$ after decomposition.
Our protocol offers substantial savings ranging from an average reduction of $3.3\times$ at $\mu=1/10$ to an average reduction of $18.2\times$ at $\mu=1/50$, demonstrating consistent trends in both online LLMs.

\subsubsection{Computation Overhead Analysis}
\label{app:memory}
To validate the practicality of our method, we investigate the user-side computation overhead in terms of memory cost and sampling cost. 

\begin{figure}[htp]
\begin{center}
{
    \includegraphics[width=0.45\columnwidth]{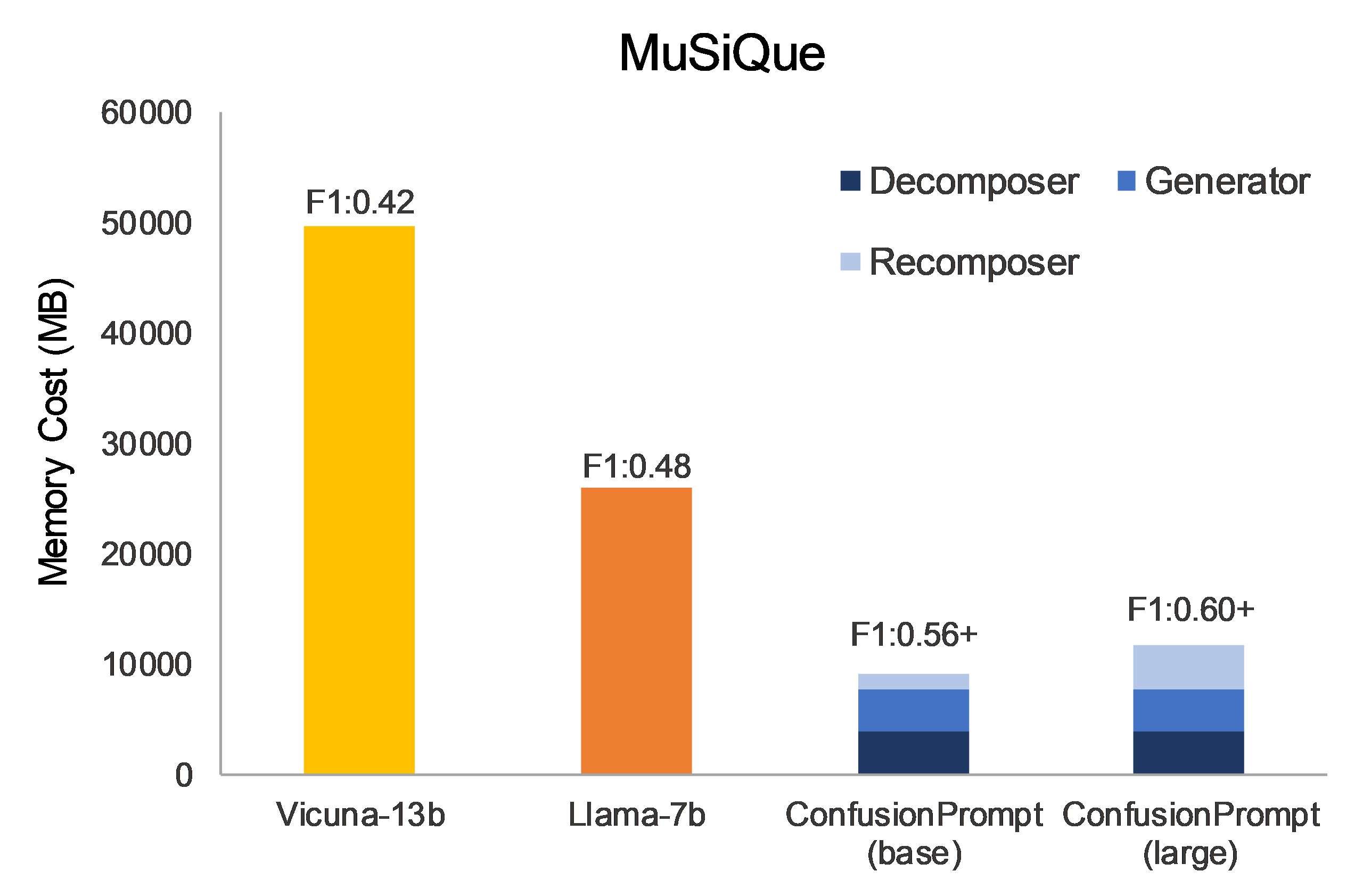}%
    \label{fig:mqmemory}
    }
\hfill
{
    \includegraphics[width=0.45\columnwidth]{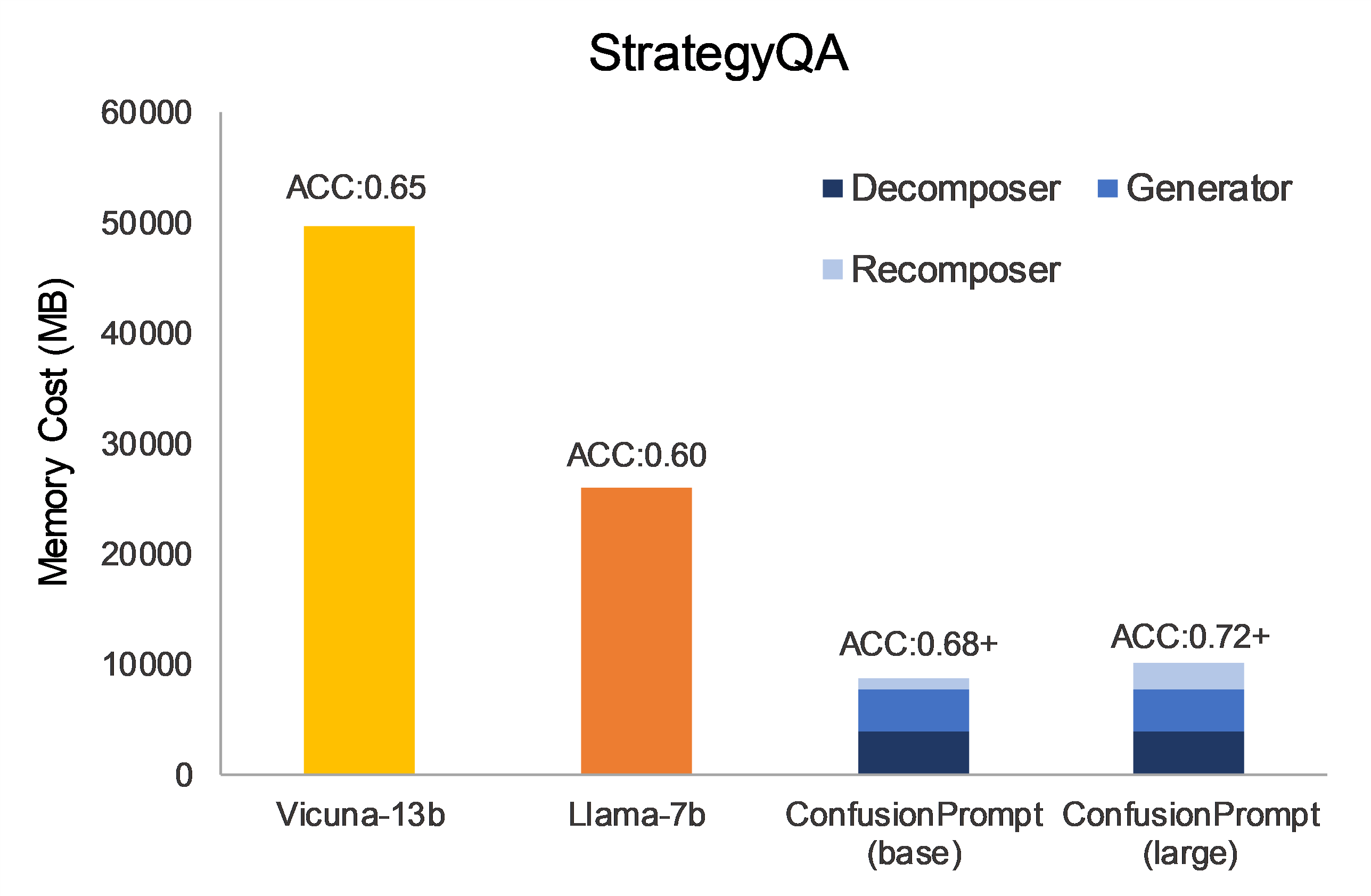}%
    \label{fig:sqamemory}
    }
\caption{User memory cost and utility on MuSiQue and StrategyQA for inference with 10 samples. ConfusionPrompt employs GPT-3.5-Turbo for online inference, with potential performance improvements using GPT-4-Turbo or GPT-4o.}
\label{fig:memory}
\end{center}
\end{figure}

\textbf{Memory Cost:} In Figure \ref{fig:memory}, we present the memory cost required for the three local models, as well as that for local open-source LLMs. For MuSiQue, ConfusionPrompt (Base) utilizes BART-Base as the recomposer, while ConfusionPrompt (Large) employs BART-Large. For StrategyQA, ConfusionPrompt (Base) uses RoBERTa-Base as the recomposer, and ConfusionPrompt (Large) uses RoBERTa-Large. 

From the results, we can see that our proposed ConfusionPrompt is significantly efficient compared to inferring large open-source language models on the user side.
Even if using the large recomposer, it requires only $21.8\%$ and $41.8\%$ memory cost on average compared to run LLaMa2-7B and Vicuna 13B, respectively, on the user side. Additionally, ConfusionPrompt provides better utility than local LLM inference with lower memory requirement. 

The empirical results also show the trade-off between memory requirement and utility for ConfusionPrompt. Using the base recomposer save the memory cost by an average of 17\% compared with the large recomposer, with inferior performances. The suitable choice of local models depends on the user's utility requirement and computational constraint.

\begin{figure}[htp]
\begin{center}
\includegraphics[width=1\linewidth]{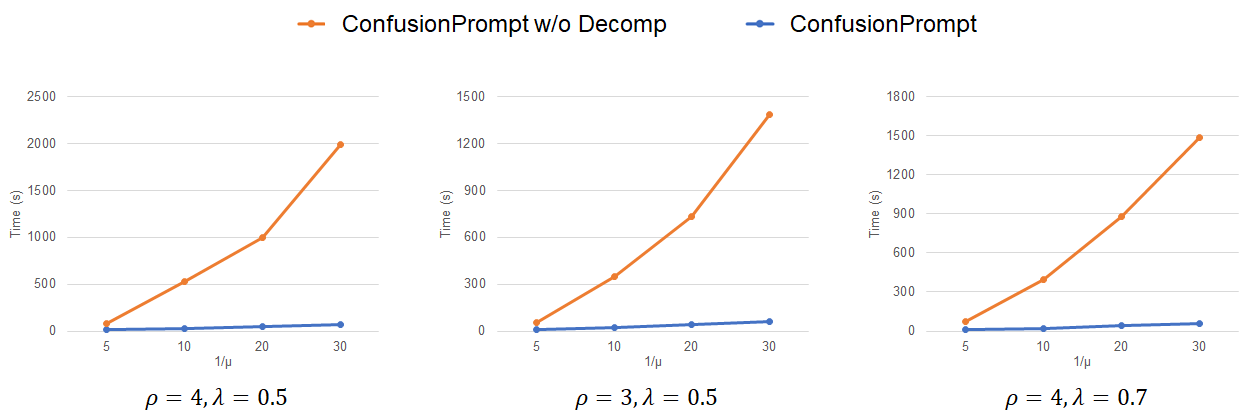}
\caption{Cost of generation in terms of time cost (seconds) for ConfusionPrompt and ConfusionPrompt w/o Decomp, under varying genuineness parameter $\rho$, similarity parameter $\lambda$, and significance parameter $\mu$.}
\label{fig:costgen}
\end{center}
\end{figure}

\textbf{Sampling Cost:} Algorithm \ref{alg:framework} suggests that the generator keeps generating fake prompts until the $(\lambda, \mu, \rho)$-privacy is satisfied. Therefore, it is crucial to consider the sampling time it takes for a generator to produce sufficient qualified pseudo-prompts. Figure \ref{fig:costgen} presents the number of generations to perform and the time cost for each generation under different levels of significance $\mu$. We consider three scenarios with varying combinations of genuineness parameter $\rho$ and similarity parameter $\lambda$.

For ConfusionPrompt, the sampling time scales approximately linear in significance requirement $1/\mu$. Our experiments indicate that under the most stringent privacy settings ($\rho=4$, $\lambda=0.5$), sampling is conducted an average of $2.5$ times per $1/\mu$. Relaxing the privacy constraints to $\rho=3$ and $\lambda=0.7$ reduces this ratio to approximately $2.1$. Note that $1/\mu$ can be treated as the minimal number of sampling times to achieve the privacy budget when the prompt contains only one private attribute.

Furthermore, we compare the generation cost for ConfusionPrompt and ConfusionPrompt w/o Decomp. We observe that decomposition reduces the sampling cost by an average of $5.88\times$ at $1/\mu=5$ and $26\times$ at $1/\mu=30$, as fewer pseudo-prompts are required to achieve the same privacy requirement.

\subsubsection{Local Model Selection}
\label{app:modelchoice}

In this section, we justify the selection of decomposer and recomposer. For decomposer, we obtain the BLEU between the generated decomposition and reference provided by the dataset. For recomposer, the models are evaluated in terms of the accuracy metrics described in Section \ref{sec:utility}, using the golden decomposition and sub-answer as training and testing inputs. 

Table \ref{tab:localcompare} presents the experiment results on MuSiQue using models from three categories (i.e., BART, T5, and Flan-T5) of varying sizes. While Flan-T5-large exhibits the best performance for both tasks, we select BART-large as the decomposition and recomposition model due to its balanced trade-off between model size and performance.

\begin{table}[htp]
\caption{Parameter size and performance on both decomposition and recomposition tasks across various models for MuSiQue.}
\label{tab:localcompare}
\vskip -0.1in
\begin{center}
\begin{sc}
\begin{small}
\scalebox{0.85}{
\begin{tabular}{lcccc}
\toprule
 & Decomposer & \multicolumn{3}{c}{Recomposer} \\
  & BLEU & F1 & RougeL & EM \\
\midrule
BART-base (139M) & 0.314 & 0.563 & 0.564 & 0.446 \\ 
BART-large (406M) & 0.402 & 0.640 & 0.641 & 0.525 \\
\hline
T5-base (223M) & 0.343 & 0.581 & 0.580 & 0.484 \\
T5-large (737M) & 0.408 & 0.646  & 0.645 & 0.554\\
\hline
Flan-T5-base (248M) & 0.367 & 0.608 & 0.609 & 0.518 \\
Flan-T5-large (783M) & \textbf{0.458} & \textbf{0.671} & \textbf{0.671} & \textbf{0.577} \\
\bottomrule
\end{tabular}
}
\end{small}
\end{sc}
\end{center}
\end{table}

\section{Discussion}
ConfusionPrompt focuses on the privacy protection for textual data during privacy-preserving LLM inference. In this section, we provide further clarifications and discussions on practical issues of our framework.

\subsection{Protection Capability of Privacy Model} 
Our $(\lambda, \mu, \rho)$-privacy model establishes requirements for fabricated prompts to mitigate privacy risks. Rather than offering a worst-case privacy guarantee such as LDP, the model quantifies the criteria that a group of prompts should satisfy. Our theoretical and empirical experiments associate the privacy parameters with attack success rate. It is important to note that the privacy model should be regarded as an enhancement over the insecure status quo, rather than a substitute for formal privacy guarantees.

Alternative privacy models could be explored to enhance privacy protection in future studies. One potential direction is to analyze the distribution of sensitive attributes conditioned on the non-sensitive components, ensuring that the set of pseudo-prompts closely approximates the ground-truth distribution.

\subsection{Trade-off between Privacy and Query Cost} 
Whereas traditional perturbation-based methods trade privacy for utility, our ConfusionPrompt trades privacy for query complexity while providing accurate responses. This approach is particularly suitable for clients with high privacy demands who can accommodate these costs, such as finance and healthcare organizations. Additionally, although query costs rise linearly with the number of tokens in API services, users of chatbot services can submit unlimited requests under a fixed subscription fee, allowing for enhanced privacy protection at a manageable cost.

The query complexity could be further reduced by considering more condensed representation of the batch of pseudo-prompt (see \ref{app:condense}). While the complexity still remains at least linearly dependent on the significance requirement $1/\mu$, future research may explore approaches to achieve a sub-linear dependence on $1/\mu$.

\subsection{Generation of Pseudo-Prompts}
The privacy protection of our framework relies on the quality of pseudo-prompts in effectively concealing sensitive attributes. In this work, we define the quality of fabricated prompts based on three key privacy considerations: the significance of attributes, the genuineness of the pseudo-query, and its similarity to the genuine prompt. Additionally, the generated prompts should maintain a consistent semantic structure for the non-sensitive components. The privacy requirement and assumption are satisfied by selecting filtered prompts from the generated candidates. Our approach employs a lightweight local generator to produce fabricated prompts; however, this may be insufficient, particularly when generating a large number of pseudo-prompts under stringent privacy requirements. To address this, we propose an alternative method in \ref{app:genquality} that leverages a powerful online LLM for pseudo-prompt generation while ensuring that private attributes remain hidden from the LLM owner.

In Section \ref{sec:privacyexp}, we present results from prompt identification attack, demonstrating the difficulty in distinguishing real prompt from pseudo-prompts under appropriate choice of privacy parameters. While more advanced discriminators may achieve higher attack success rates, developing stronger attackers is left for future work. This may involve using LLMs as discriminators and integrating domain knowledge. To further enhance privacy protection, future improvements could: (1) enhance quality assessment by incorporating additional privacy requirement, such as the diversity of pseudo-prompts, to strengthen privacy protection; (2) refine the evaluation of genuineness and similarity in experiments by considering metrics beyond fluency scores and embedding cosine similarity; (3) train the generator using adversarial training techniques and employ a set of sophisticated discriminators to select qualified pseudo-prompts, thereby improving the robustness of pseudo-prompt generation.

\subsection{User Control over Sensitive Data}
For privacy protection, we implement the decomposer and generator locally on the user side, granting users complete control over their sensitive data. This setup allows users to customize pseudo-prompt generation based on their privacy and ethical requirements, domain expertise, and computational resources. In particular:
\begin{itemize}
    \item Users can freely set the privacy parameters $(\lambda, \mu, \rho)$. For stringent privacy needs, users might: (1) assign a lower value to the significance parameter $\mu$, resulting in a larger set of pseudo-prompts; (2) set a lower similarity parameter $\lambda$ and a higher genuineness parameter $\rho$ by adjusting the filtering threshold.
    \item During pseudo-prompt generation, users can filter prompts to achieve the requirement of $\lambda$ and $\rho$ through: (1) utilizing a local similarity and discrimination model (see Appendix \ref{app:trainsimdis}), ensuring intermediate results remain on the user's device for enhanced privacy; (2) employing another remote LLM to assess pseudo-prompt quality, offloading computational tasks to a server, which maintains privacy if the two LLM servers do not collaborate. Additionally, users can manually review and modify prompt groups before sending them to the server. 
\end{itemize}

\section{Conclusion}
This paper proposes a private inference framework for online LLMs, termed ConfusionPrompt.
We deploy three local models on the user side: (i) decomposer that maps an original prompt to a sequence of sub-prompts, (ii) generator that produces pseudo-prompts by replacing private attributes in the genuine sub-prompts, (iii) recomposer that maps the decomposed prompt-response pairs from the cloud service to the final response.
Such design endows our framework with advantages over previous protocols that: (i) it can be seamlessly integrated with existing black-box LLMs where the service providers have no need to modify their already-built framework, and (ii) it achieves better privacy-utility trade-off than existing LDP-based privatization methods.
We develop a $(\lambda, \mu, \rho)$-privacy model to formulate the requirement for a privacy-preserving group of prompts, and accordingly  provide a complexity analysis to demonstrate the benefits of the decomposition module.
Experiments show that our ConfusionPrompt outperforms local inference with open-source models, while reducing memory consumption by at least 62\% compared to open-source LLMs. Furthermore, it achieves over 42\% higher utility than LDP-based methods.

\bibliographystyle{IEEEtran}
\bibliography{reference}
\begin{appendix}
\subsection{Proof of Theorem \ref{theo:complexsingle} and \ref{theo:complexdecomp}}
\label{app:complexproof}
We begin with the proof for Theorem \ref{theo:complexsingle} as follows:

\begin{proof}
    For the prompt with single paragraph, the significance can by represented as:
\begin{equation}
\begin{gathered}
\label{eq:singlesig}
    \mathrm{Sig}(U(p_0), \boldsymbol{P}) =
    \max_{u_k}  
    \max_{\mathcal{V}} 
    \frac{\sum_{i=1}^n \bar{H}(\mathcal{V}, p_0, p_i)}{\sum_{i=1}^n \bar{H}(\mathcal{V}\backslash u_k, p_0, p_i)} \\
    {\rm s.t.} \ u_k \in \mathcal{V}\subseteq U(p_0)), 
\end{gathered}
\end{equation}
where $p_0$ is the genuine prompt, and $p_i$ is the $i^{\mathrm{th}}$ prompt in the prompt group, and:
\begin{equation}
\bar{H}(\mathcal{V}, p_0, p_i) =\left\{
\begin{array}{ll}
1 & \mathrm{Corr}\left(u,p_0, p_i\right)=u\ \forall u \in \mathcal{V} \\
& \mathrm{or}\ \mathcal{V}=\emptyset\\
0 & \rm{otherwise}
\end{array}.
\right.
\end{equation}
For $\mathrm{Sig}(U(p_0), \boldsymbol{P}) = \mu$, it is obvious that there should be at least $1/\mu$ distinct values, including genuine and fake values, assigned to each attribute. Otherwise suppose $u_k$ have less than $1/\mu$ distinct values, we would have:
\begin{equation}
   \frac{\sum_{j=1}^{|\boldsymbol{P}|}H(u_k, p_0, p_j)}{|\boldsymbol{P}|} > \mu,
\end{equation}
which violates the expression \ref{eq:singlesig}.

Next, we claim that each possible combination of attribute set values must appear at least once within the prompt group. To proof by contradiction, let's suppose that there is one combination missing in the group while the others appear once. Then we can consider two cases:

(a) The missing combination does not contain any true values. Then the significance of other single attribute must be larger than $\mu$. This is because for a set containing all combinations, the proportion of the occurrence for each attribute is exactly $\mu$. Therefore, such removal would make their significance smaller.

(b) The missing combination contains a subset of true values $\mathcal{V}\subseteq U(p_0)$. Then conditioned on the subset $\mathcal{V}$, the proportion of any other single attribute would be over $\mu$, as their original proportion is $\mu$. Therefore, for any $u_k \notin \mathcal{V}$, it holds that:
\begin{equation}
     \frac{\sum_{i=1}^n \bar{H}(\mathcal{V} \cup u_k, p_0, p_i)}{\sum_{i=1}^n \bar{H}(\mathcal{V}, p_0, p_i)} > \mu,
\end{equation}
which violates the expression \ref{eq:singlesig}.

To make each combination occur at least once, we should have at least $(1/\mu)^{|U(p_0)|}$ prompts in the group.
\end{proof}
Following that, we can proceed to the proof for Theorem \ref{theo:complexdecomp}:
\begin{proof}
   Let $\mu=\mathrm{Sig}(U(p_0), \boldsymbol{P})$ denotes the significance. According to Theorem \ref{theo:complexsingle}, for each sub-query $p_{0h}\in \boldsymbol{p}_0$, we should have at least $(1/\mu)^{\|U(p_{0h})\|}$ prompts to satisfy:
\begin{equation}
\begin{gathered}
    \max_{u_k}  
    \max_{\mathcal{V}} 
    \frac{\sum_{i=1}^n \bar{H}(\mathcal{V}, p_{0h}, p_{ih})}{\sum_{i=1}^n \bar{H}(\mathcal{V}\backslash u_k, p_{0h}, p_{ih})} \leq \mu \\
    {\rm s.t.} \ u_k \in \mathcal{V}\subseteq U(p_{0h}).
\end{gathered}
\end{equation}
Then the proof completes by summing up the number of prompts over all sub-queries.
\end{proof}

\subsection{Bound on Inference Attack}
\label{app:attacksig}
\subsubsection{Inference from Combination Pattern}
We consider the scenario where an attacker attempts to identify the target attribute by prior knowledge of other attributes and the combination pattern of attribute values. The following theorem states that an curious server can do no better than random guessing if the combination of values are uniformly distributed:
\begin{lemma}
Suppose the attacker is guessing the attributes from attribute combination pattern in $\boldsymbol{P}$. The attacker can do no more better than random guessing if the following condition is satisfied:
\begin{itemize}
    \item Each combination of attribute set has the same occurrence times in the prompt groups $\boldsymbol{P}$, i.e., the combinations are uniformly distributed.
\end{itemize}
\end{lemma}
From the proof of Theorem \ref{theo:complexsingle} and \ref{theo:complexdecomp}, \emph{we can construct a uniformly distributed prompt group under the optimal number of prompts.} Therefore, we can turn to the success rate under a random guessing attack, which is bounded by Lemma \ref{lem:attackbound}.

\begin{lemma}
\label{lem:attackbound}
Given a group of prompts $\boldsymbol{P}=\{\boldsymbol{p}_0, \boldsymbol{p}_1, ..., \boldsymbol{p}_n\}$ with $(\lambda, \mu, \rho)$-privacy, suppose that the attacker has prior knowledge about the attribute subset $\mathcal{V}\subseteq U(\boldsymbol{p}_0)$. By random guessing, the probability of correctly identifying any target attribute $u\in U(\boldsymbol{p}_0) \land u\notin \mathcal{V}$ is upper bounded by $\mu$.
\end{lemma}
\begin{proof}
    According to Definition \ref{def:sigset}, it holds that:
\[
    \frac{\sum_{j=1}^n \bar{H}(\mathcal{V}, p_{0h}, p_{jh})}{\sum_{j=1}^n \bar{H}(\mathcal{V}\backslash u, p_{0h}, p_{jh})} \leq \mu
\]
for any $h$ and $u$. In other words, the proportion of any attribute $u$ conditioned on the prior knowledge attribute set $\mathcal{V}$ is upper bounded by $\mu$. Therefore, by random guessing, the probability of correctly identifying any target attribute $u$ is upper bounded by $\mu$.
\end{proof}

An interpretation of the above lemmas is that \emph{under $(\lambda, \mu, \rho)$-privacy, we can always construct a group of prompts with optimal complexity, where the probability of correctly identifying any target attribute is upper bounded by $\mu$, if the attacker utilizes the combination pattern of attributes (ignoring the semantic meaning)}.

\subsubsection{Inference from Contextual Information}
The above analysis ignores the semantic information in pseudo-prompts. We further consider the scenario where an attacker attempts to identify the target attribute using optimal reconstruction attacks that leverage the semantic information in pseudo-prompts \cite{tong2025vulnerability}. Under such an attack, the attacker: (i) selects the prompt most likely to be the true prompt; and (ii) infers the underlying attribute based on the selected prompt. We still assume that the attacker has prior knowledge about the attribute subset $\mathcal{V}\subseteq U(\boldsymbol{p}_0)$. The optimal prompt selection under Bayes' rule is given by:
\[
\begin{gathered}
q_{\text{pid}}(\boldsymbol{P},\mathcal{V}) = \arg\max_{\boldsymbol{p}\in \boldsymbol{P}}\Pr(\boldsymbol{p}\mid \boldsymbol{P},\mathcal{V})\\
=\arg\max_{\boldsymbol{p}\in \boldsymbol{P}}\Pr(\boldsymbol{P}\mid \boldsymbol{p},\mathcal{V})\Pr(\boldsymbol{p}\mid \mathcal{V}).
\end{gathered}
\]

Given the selected prompt $\boldsymbol{p}^*$, the probability to identify the target attribute $u_k$ is defined through a similarity-based function:
\[
\begin{gathered}
\Pr(q_{\text{attr}}(\boldsymbol{p}^* \mid \boldsymbol{P}))=
g_{\text{sim}}\left(u_k, \boldsymbol{p}^*\right)\\
\leq\left\{
\begin{array}{ll}
g_1\left(u_k, \boldsymbol{p}^*\right) & \mu_k\in U(\boldsymbol{p}^*) \\
g_\lambda\left(u_k, \boldsymbol{p}^*\right) & \mu_k\notin U(\boldsymbol{p}^*)
\end{array},
\right.
\end{gathered}
\]
where $g_{\text{sim}}(\cdot)$ denotes the probability given the similarity between $u_k$ and the corresponding attribute in $\boldsymbol{p}^*$, and the inequality follows from the fact that $g_{\text{sim}}(\cdot)$ is increasing in the similarity.

Under this Bayes rule, the success probability is upper bounded by:
\[
\begin{gathered}
\mathrm{ASR}^\star = \Pr(g_{\text{sim}}\left(u_k, \boldsymbol{p}^*\right)=u_k) \\
\leq \Pr\left(u_k\in U(\boldsymbol{p}^*)\right) \left(g_1\left(u_k, \boldsymbol{p}^*\right)-g_\lambda\left(u_k, \boldsymbol{p}^*\right)\right) + g_\lambda\left(u_k, \boldsymbol{p}^*\right)
\end{gathered}
\]
The term $\Pr\left(\mu_k\in U(\boldsymbol{p}^*)\right)$ is bounded by:
\[
\begin{gathered}
\Pr\left(u_k\in U(\boldsymbol{p}^*)\right) = \mathbb{E}_{\boldsymbol{P}}[\max_{\boldsymbol{p}\in \boldsymbol{P}}\Pr(\boldsymbol{p}\mid \boldsymbol{P},\mathcal{V})]\\
\leq \mathbb{E}_{\boldsymbol{P}}[1-(1/u-1)\min_{\boldsymbol{p}\in \boldsymbol{P}}\Pr(\boldsymbol{p}\mid \boldsymbol{P},\mathcal{V})].
\end{gathered}
\]
Since $\Pr(\boldsymbol{p}\mid \boldsymbol{P},\mathcal{V})$ is correlated with the genuineness score $\rho$, it can be modeled with a function $g'_{Genu}(\boldsymbol{p}, \boldsymbol{P})$. As it increases with the genuineness score, we have:
\[
\begin{gathered}
\Pr\left(u_k\in U(\boldsymbol{p}^*)\right)
\leq \mathbb{E}_{\boldsymbol{P}}[1-(1/u-1)g'_{\rho}(\boldsymbol{p}, \boldsymbol{P})].
\end{gathered}
\]
Combining all together, the attribute inference attack is upper bounded by:
\[
\begin{gathered}
\mathrm{ASR}^\star\leq g_\lambda\left(u_k, \boldsymbol{p}^*\right)
\\+ \mathbb{E}_{\boldsymbol{P}}[1-(1/u-1)g'_{\rho}(\boldsymbol{p}, \boldsymbol{P})] \left(g_1\left(u_k, \boldsymbol{p}^*\right)-g_\lambda\left(u_k, \boldsymbol{p}^*\right)\right).
\end{gathered}
\]
\subsection{Specifications of Dataset}
\label{app:cat}

StrategyQA and MuSiQue are multi-hop reasoning questions that require multiple reasoning steps to arrive at the correct answer. MMLU is a widely adopted benchmark designed to evaluate an LLM's knowledge and problem solving abilities. 

\begin{figure}[htp]
\begin{center}
{
    \includegraphics[width=0.30\linewidth]{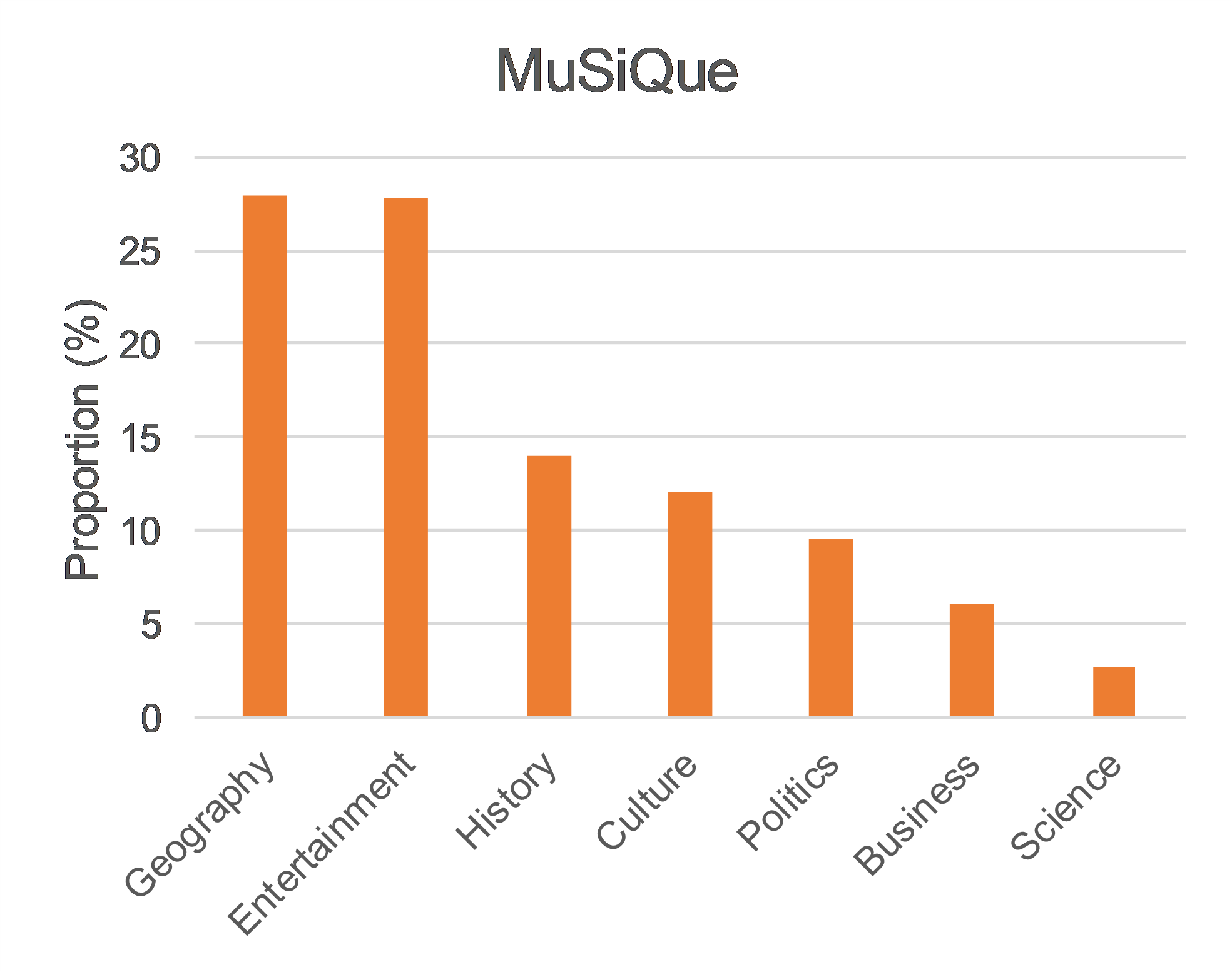}%
    \label{fig:mqcat}
    }
{
    \includegraphics[width=0.30\linewidth]{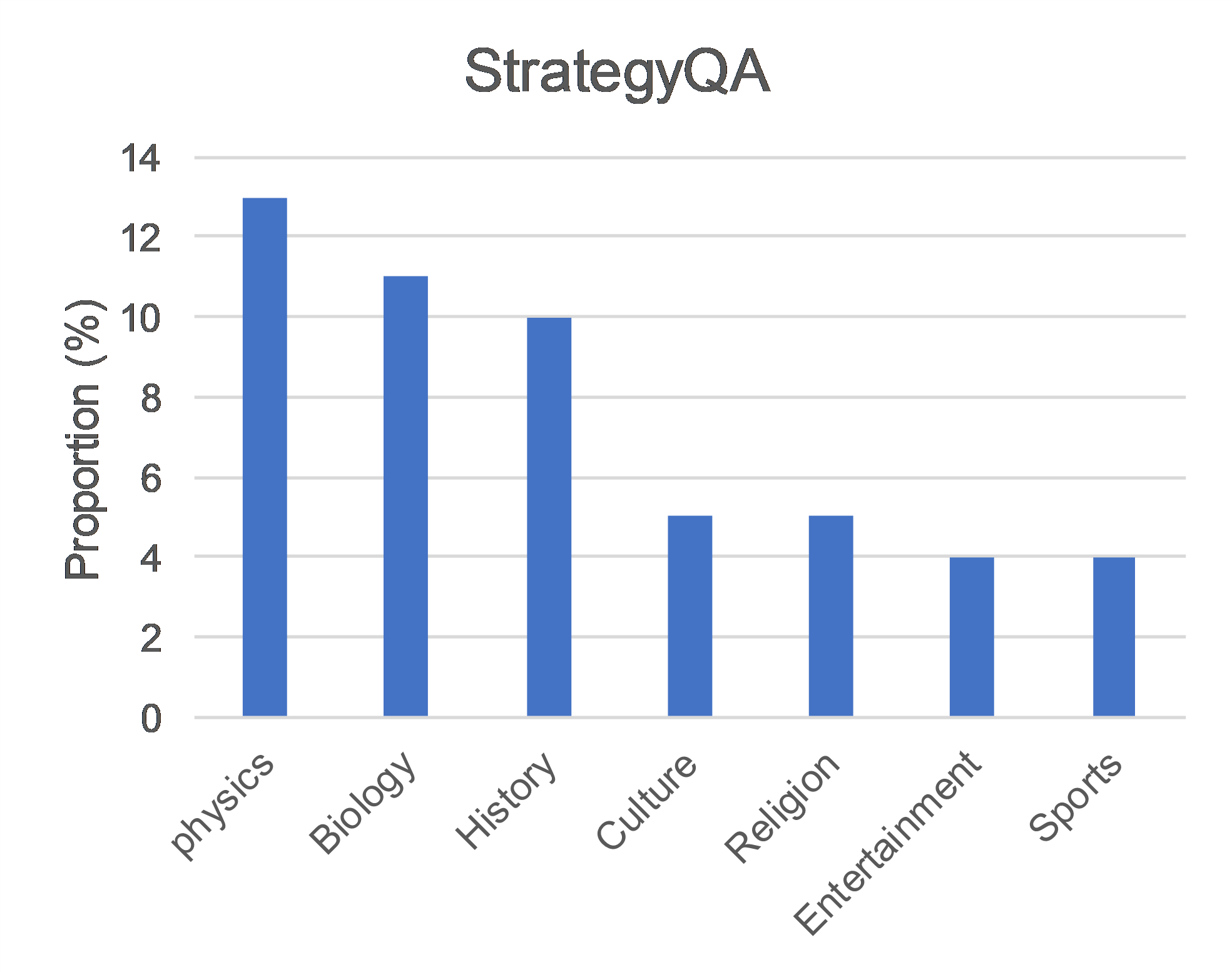}%
    \label{fig:sqacat}
    }
{
    \includegraphics[width=0.30\linewidth]{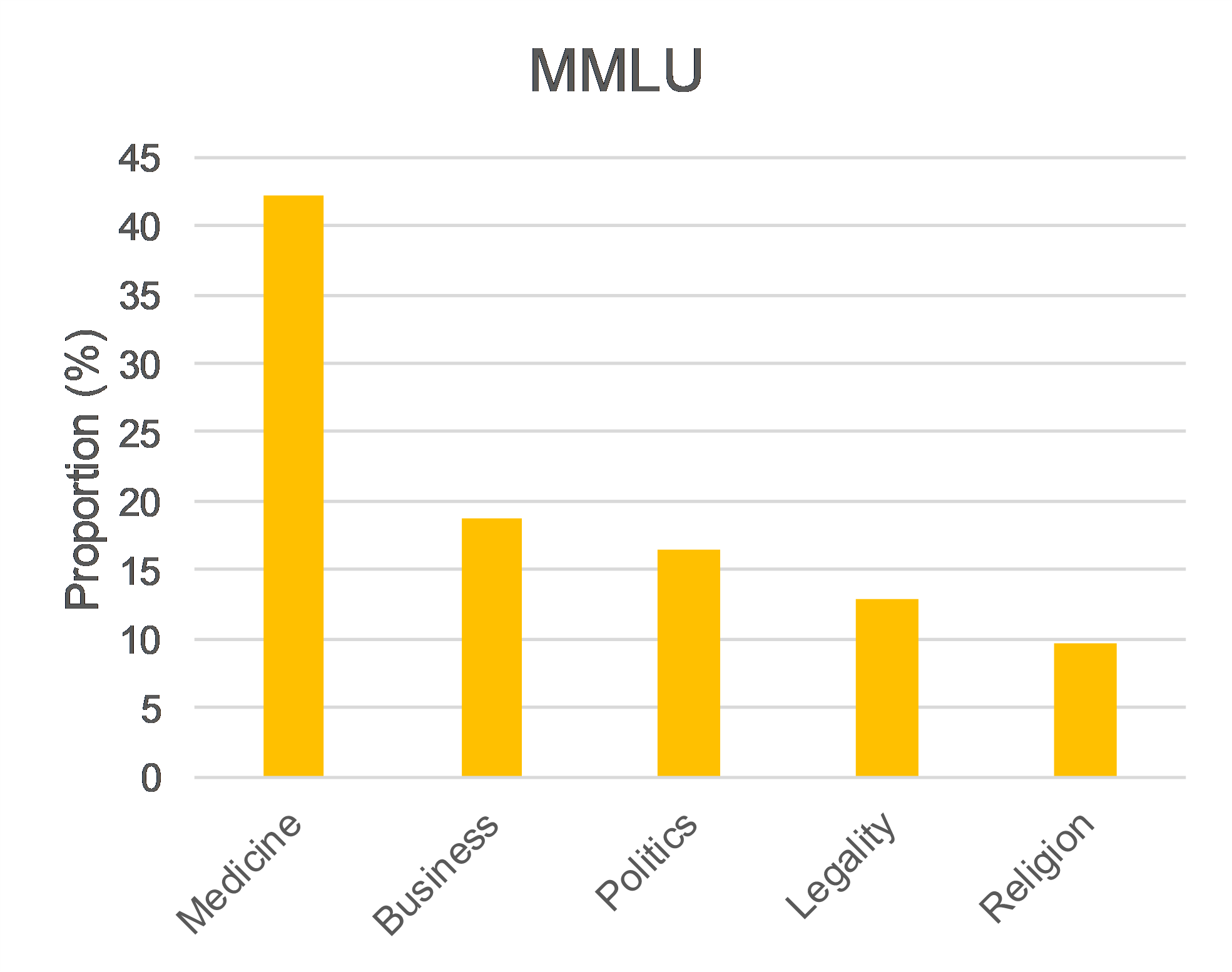}%
    \label{fig:mmlucat}
    }
\caption{Proportion of subjects for StrategyQA, MuSiQue, and MMLU datasets.}
\label{fig:catmultihop}
\end{center}
\end{figure}

Figure \ref{fig:catmultihop} presents the distribution of subjects for the three datasets. For StrategyQA and MuSiQue, topics such as religion, business, and politics have higher tendency to contain sensitive information among their subjects. To complement the two multi-hop datasets, we sample a subset of privacy-related questions from the MMLU dataset, containing questions from five sensitive areas: business, legality, politics, medicine, and religion.

\subsection{Construction of Private Attributes}
\label{app:attrconst}

During step 1, we label the private attributes according to the following guidelines:
\begin{itemize}
    \item Identify the key components for a query, including but not limited to proper nouns, phrases, verbs, and adjectives.
    \item Ensure that there is no overlapping information between different private attributes within the same query.
    \item Each private attribute should be as succinct as possible. For example, instead of labeling "spouse of the Green performer" as a single attribute, it is preferable to split it into "spouse" and "Green performer" as separate attributes.
\end{itemize}
The third point is to ensure privacy protection even when an attacker possesses prior knowledge of certain attributes.
Table \ref{tab:privateattr} provides several examples of private attributes.

\begin{table}[htp]
\caption{Examples of private attributes in StrategyQA and MuSiQue.}
\label{tab:privateattr}
\centering
\begin{tabularx}{\columnwidth}{@{} X X @{}}
\toprule
\multicolumn{1}{c}{\textsc{Query}} & \multicolumn{1}{c}{\textsc{Private Attributes}}\\
\midrule
Are \textcolor{blue}{blue lips normal}? & blue lips, normal\\
\midrule
Could \textcolor{blue}{ten gallons} of \textcolor{blue}{seawater crush} a \textcolor{blue}{six year old}? & seawater crush, ten gallons, six year old\\
\midrule
Who is the \textcolor{blue}{spouse} of the \textcolor{blue}{Green performer}? & spouse, Green performer\\
\midrule
What \textcolor{blue}{instrument} is \textcolor{blue}{played} by the person from \textcolor{blue}{The Blackout All-Stars}? & play instrument, The Blackout All-Stars\\
\midrule
What is the \textcolor{blue}{capital of the county} that \textcolor{blue}{Pine Springs} is located in? & capital of the county, Pine Springs\\
\bottomrule
\end{tabularx}
\vskip -0.1in
\end{table}

In step 2, we evaluate the generative quality of various LLMs on five models: i) GPT-4-Turbo with few-shot examples, ii) GPT-3.5-Turbo finetuned with 100 samples, iii) BART-large \cite{lewis2020bart}, iv) T5-large \cite{raffel2020exploring}, and v) Flan-T5-Xlarge \cite{chung2024scaling}. The latter three open-source models are finetuned with 1000 samples.

Table \ref{tab:attrextract} presents the performance on a human-labeled test dataset. It can be observed that the finetuned version of GPT-3.5-turbo gives the best result, and thus it's used as the attribute extraction model throughout our experiment.

\begin{table}[htp]
\caption{Comparison of various attribute extraction models.}
\label{tab:attrextract}
\begin{center}
\scalebox{0.85}{
\begin{sc}
\begin{tabular}{lcccccc}
\toprule
\multirow{2}{*}{Method} & \multicolumn{3}{c}{StrategyQA} & \multicolumn{3}{c}{MuSiQue} \\
& F1 & roughL & EM & F1 & roughL & EM \\
\midrule
GPT-4-Turbo & 0.689 & 0.748 & 0.080 & 0.811 & 0.762 & 0.234 \\
GPT-3.5-Turbo & \textbf{0.803} &  \textbf{0.881} & \textbf{0.421}  &  \textbf{0.846} & \textbf{0.789} & \textbf{0.333} \\
BART-large & 0.713 & 0.690  & 0.227  & 0.664 & 0.634 & 0.118 \\
T5-large & 0.603 & 0.595 & 0.102 & 0.381 & 0.383 & 0.010\\
Flan-T5-large & 0.669 & 0.683 & 0.243 & 0.380 & 0.383 & 0.026 \\
\bottomrule
\end{tabular}
\end{sc}
}
\end{center}
\end{table}
\subsection{Training of local models}
\label{app:localtrain}
In this section, we describe the training of decomposer, generator, and recomposer, respectively.

\textbf{Training of decomposer:}
The decomposition model is finetuned on a pretrained BART model \cite{lewis2020bart} with 406M parameters.
In the first stage, we utilize the decomposition samples from the three datasets as demonstration data.
In the second stage, we evaluate the response through the following steps: (i) extract the sub-answers using gpt-4-turbo and obtain the final answers using a recomposition model to judge whether the decomposition returns a correct answer;
(ii) compute the complexity measured by the number of required pseudo-prompts under $\mu=10$. 

During step (i), to address the recomposer's limitations in determining the correctness of the final answer, we train a robust recomposition model by fine-tuning LLaMa2-7B. Note that during inference, we adopt a much smaller recomposition model for efficiency.

\textbf{Training of generator:}
The generator is finetuned on the BART-large model.
To collect demonstration data for the first stage, we prompt GPT-4-turbo to generate multiple pseudo-prompts given the raw prompt and its private attributes using the following template.
\begin{small}
\begin{mdframed}[backgroundcolor=black!5]
\ \ \ Please replace the phrases \{\textcolor{red}{attributes}\} in the each of the following sentences, such that each sentence is fluent and reasonable, and the alternative phrases have irrelevant meaning as \{\textcolor{red}{attributes}\}.

Please return \{\textcolor{red}{\# of replaces}\} replacements for each sentence.

Strictly respond in the form of JSON with the following format: \{"S1": ["replacement 1", "replacement 2",...], "S2": ["replacement 1", "replacement 2",...]\}.

Sentences: \{\textcolor{red}{dictionary of sentences}\}
\end{mdframed}
\end{small}

In the second, we employ two models to evaluate the semantic similarity and genuineness, including a local similarity evaluation model and discriminator.
For similarity evaluation model, we adopt a finetuned version of MiniLM-6L model \cite{reimers2019sentence} to extract the embedding of each private attribute. The semantic relevance between a pair of attributes is given by the cosine similarity between their embeddings.
For discriminator, we leverage GPT-4o to construct the training dataset for training a local discriminator. As the genuineness is closely related to the sentence's fluency, we instruct GPT-4o to evaluate the fluency for each sentence \cite{kanumolu2023unsupervised} (see detailed instruction in \ref{app:trainsimdis}).

\textbf{Training of recomposer:}
The recomposition model is finetuned on a pretrained RoBERTa \cite{liu2019roberta} and BART-large \cite{lewis2020bart} model, respectively, for strategyQA and MuSiQue. We utilize the sub-prompts \& sub-responses given in StrategyQA and MuSiQue datasets as well as the final responses to train a recomposition model. For MMLU, we construct the sub-queries and sub-answers using GPT-4o. To improve the performance of recomposer, we pretrain the model on two additional datasets, SQuAD \cite{rajpurkar2016squad} and DROP \cite{dua2019drop}.

\subsection{Semantic Similarity Model and Discriminator}
\label{app:trainsimdis}

The comparison data collection for the generator involves a local similarity evaluation model and discriminator.

\textbf{Similarity evaluation model:} We adopt a finetuned version of MiniLM-6L model \cite{reimers2019sentence} to extract the embedding of each private attribute. The semantic relevance between a pair of attributes is given by the cosine similarity between their embeddings.

\textbf{Discriminator:} We leverage GPT-4 to construct the training dataset for training a local discriminator. As the genuineness is closely related to the sentence's fluency, we prompt GPT-4 to evaluate the fluency for each sentence with the following template \cite{kanumolu2023unsupervised}:
\begin{small}
\begin{mdframed}[backgroundcolor=black!5]
\ \ \ Given multiple sentences, use the scoring rules below to score each sentence's fluency on a scale of 1 to 4:

1. Score 1: Incomprehensible. Inarticulate/ non-fluent sentence.

2. Score 2: Low Quality. Partially fluent sentence: (a) only half of the sentence is fluent or (b) more than 1 missing words or (c) more than 1 misspelt words or d) contains individual fluent word-groups with missing coherence between them.

3. Score 3: Moderate. Sentence is predominantly fluent but contains either (a) misspelt word or (b) missing word or (c) multiple occurrence of a word. 

4. Score 4: Perfect. Perfectly fluent sentence without any syntactic or grammatical error.

Strictly respond in the form of JSON with the following format: 
\{"S1": the score, "S2": the score\}.

Sentences: \{\textcolor{red}{dictionary of sentences}\}
\end{mdframed}
\end{small}

On obtaining 4000 training and 700 validation samples, we finetune a Bert-base (110M parameters) to train a local discriminator. 

\subsection{Results on Prompt Identification Attack}
\label{app:attack}
\begin{figure}[htp]
\begin{center}
{
    \includegraphics[width=0.45\columnwidth]{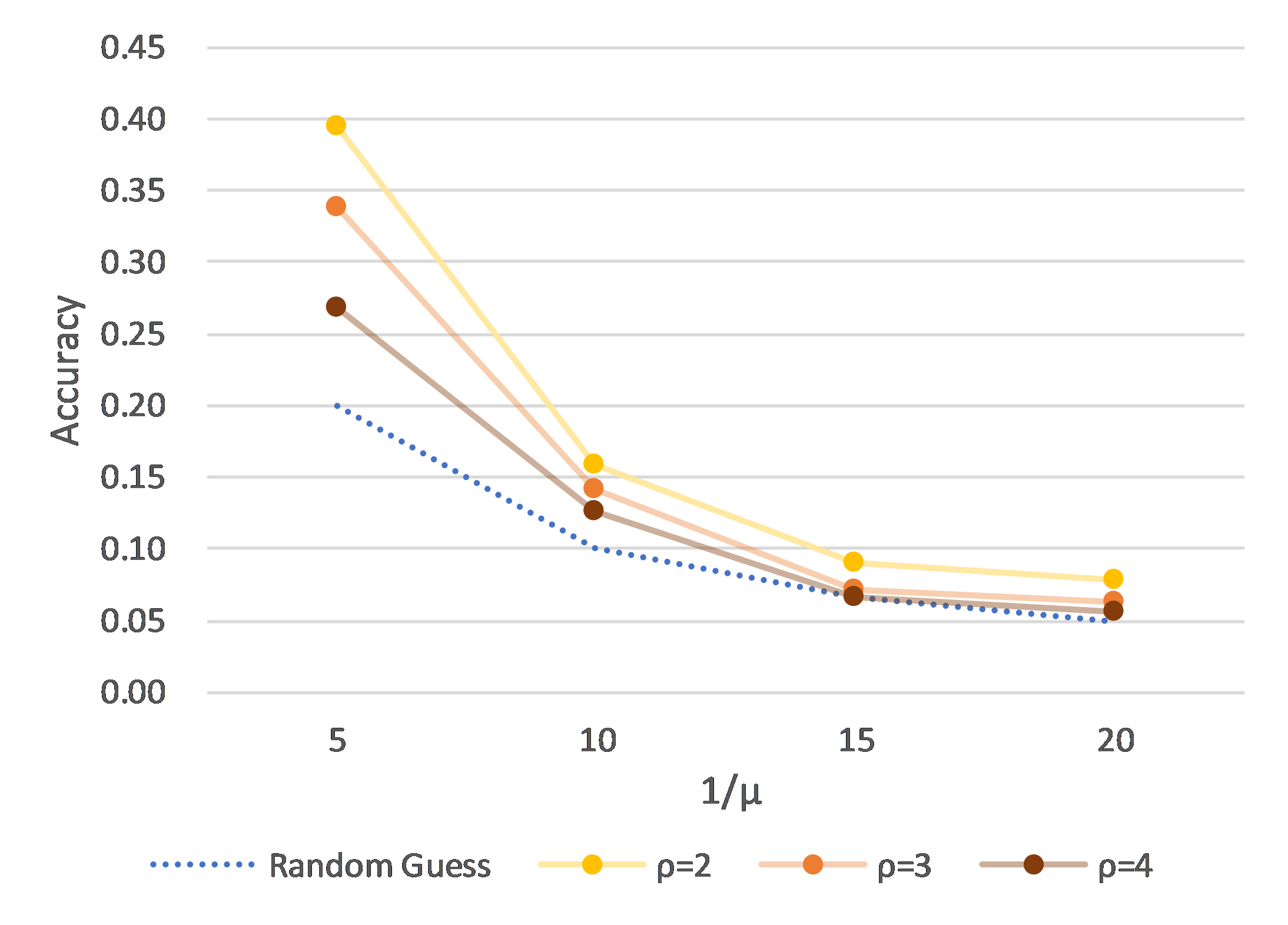}%
    \label{fig:phomu}
    }
\hfill
{
    \includegraphics[width=0.45\columnwidth]{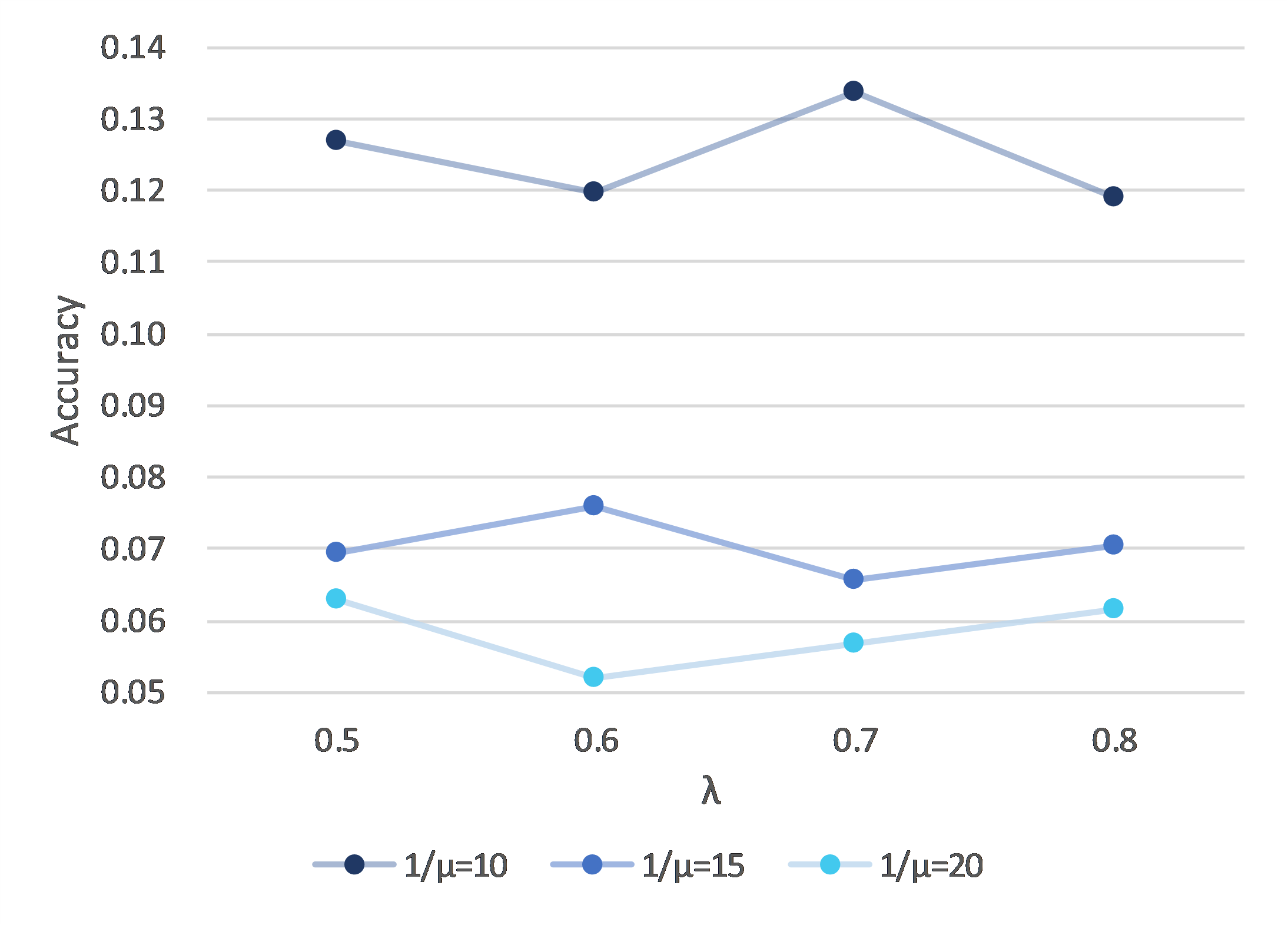}%
    \label{fig:mulambda}
    }
\caption{Prompt identification attack accuracy under various combinations of significance parameter $\mu$, genuineness parameter $\rho$, and similarity parameter $\lambda$.}
\label{fig:piattackapp}
\end{center}
\end{figure}

Figure \ref{fig:piattackapp} presents additional results of prompt identification attack under varying combinations of privacy parameters. The results suggest that the attack accuracies decrease as the increase in $1/u$ and $\rho$. Furthermore, $\lambda$ does not exhibit explicit relationship with the attack success rate.


\subsection{Condense Representation of Pseudo-Prompts}
\label{app:condense}
We propose a query-efficient approach to transmit the group of pseudo-prompts. Instead of sending $n$ prompts individually, a user could share the non-sensitive part across all prompts. For example, given the original question "Are psychiatric patients welcome to join the United States Air Force?" and the private attributes "psychiatric patients" and "United States Air Force", a condensed representation could be as follows:

\begin{small}
\begin{mdframed}[backgroundcolor=black!5]
\ \ \ Given the question: Are \#ATTR1 welcome to join the \#ATTR2? Please provide the answers respectively by replacing the \#ATTR with the \{\textcolor{red}{n}\} combinations:

1. \#ATTR1: \{\textcolor{red}{attribute 1.1}\},  \#ATTR2: \{\textcolor{red}{attribute 2.1}\}

2. \#ATTR1: \{\textcolor{red}{attribute 1.2}\},  \#ATTR2: \{\textcolor{red}{attribute 2.2}\}

...

\{\textcolor{red}{n}\}. \#ATTR1: \{\textcolor{red}{attribute 1.n}\},  \#ATTR2: \{\textcolor{red}{attribute 2.n}\}
\end{mdframed}
\end{small}

Let $S$ and $S_{target}$ represent the sets of tokens for the raw query and target attributes, respectively. The condensed representation reduces the number of query tokens from $O(nS)$ to $O(nS_{target})$, where $n$ denotes the number of prompts required to satisfy the privacy requirement and is at least linear in $1/\mu$.

\subsection{Generation Quality}
\label{app:genquality}
The local lightweight generator might be insufficient to produce a larger number of diverse pseudo-prompts. For more reliable generation, the user could leverage a powerful online LLM, such as GPT-4o, to produce pseudo-prompts of higher qualities without reveal their sensitive attributes.

Given a private query, a user can replace sensitive information with \#MASK and request an online LLM to generate a coherent prompt by filling in the \#MASK, following the template below:

\begin{small}
\begin{mdframed}[backgroundcolor=black!5]
\ \ \  Given a sentence, please fill in the mask, and generate \{\textcolor{red}{n}\} distinct coherent sentences. 

Remember to return in the format of ["sentence 1", "sentence 2", ...., "sentence \{\textcolor{red}{n}\}"].

Sentence: \{\textcolor{red}{dictionary of sentences}\}

Please return only a list of ["sentence 1", "sentence 2", ...., "sentence \{\textcolor{red}{n}\}"].
\end{mdframed}
\end{small}

We compare the quality of pseudo-prompts generated by the local and online methods. For both approaches, we employ the sampling strategy outlined in Line 3 of Algorithm \ref{alg:framework}, using a fluency threshold $\rho=3$ and similarity threshold $\lambda=0.5$. The results indicate that while both methods achieve similar cosine similarity scores, leveraging the online LLM enhances the fluency score by 15\% on average. Therefore, users with insufficient computation power are recommended to adopt the online generation methods for large-scale pseudo-prompt generation.

\begin{table}[htp]
\caption{Fluency score and cosine similarity of pseudo-prompts under varying $\mu$. Local generator employs the BART-Large model, and online generator utilizes the GPT-4o.}
\label{tab:quality}
\begin{center}
\begin{sc}
\begin{small}
\scalebox{0.98}{
\begin{tabular}{llcccc}
\toprule
\multicolumn{2}{c}{$1/\mu$} & 5 & 10 & 15 & 20 \\
\midrule
\multirow{2}{0.15\columnwidth}{Fluency Score} & Local & 3.54\tiny$\pm$0.88  & 3.42\tiny$\pm$0.92 & 3.46\tiny$\pm$0.88 & 3.37\tiny$\pm$1.05\\ 
 & Online & 3.95\tiny$\pm$0.29 & 3.98\tiny$\pm$0.17 & 3.98\tiny$\pm$0.19 & 3.95\tiny$\pm$0.33\\
\hline
\multirow{2}{0.15\columnwidth}{Cosine Similarity} & Local & 0.25\tiny$\pm$0.08 & 0.33\tiny$\pm$0.07 & 0.30\tiny$\pm$0.09 & 0.27\tiny$\pm$0.08 \\ 
 & Online & 0.27\tiny$\pm$0.08 & 0.30\tiny$\pm$0.09 & 0.29\tiny$\pm$0.07 & 0.29\tiny$\pm$0.09\\
\bottomrule
\end{tabular}
}
\end{small}
\end{sc}
\end{center}
\end{table}

\end{appendix}

\end{document}